\documentclass[10pt,aps,prc,floatfix,twocolumn,superscriptaddress,nofootinbib]{revtex4-2}

\usepackage[dvipsnames]{xcolor}
\usepackage{amsfonts,amsmath,amssymb,bm}
\usepackage{graphicx}
\usepackage[utf8]{inputenc}
\usepackage{xspace}
\usepackage{dcolumn}
\usepackage{multirow}
\newcolumntype{d}[1]{D{.}{.}{#1}}
\usepackage{cellspace}
\usepackage{isotope}
\usepackage{xparse}
\usepackage{physics}
\usepackage{dsfont}
\usepackage[pdfpagelabels, pdfencoding=auto, psdextra]{hyperref}
\hypersetup{%
 pdfsubject=Paper,
 pdfkeywords={nuclear physics} {Bayesian Regression} {chiral EFT} {nuclear matter} {nuclear saturation} {density functional theory}
 unicode = true,
 breaklinks = true,
 colorlinks = true,
 linkcolor = blue,
 menucolor = blue,
 citecolor = blue,
 urlcolor = blue
}

\graphicspath{{./figures/}}

\newcommand{\eg}{\textit{e.g.}\xspace}
\newcommand{\ie}{\textit{i.e.}\xspace}
\newcommand{\etal}{\textit{et~al.}\xspace}

\newcommand{\fmi}{\, \text{fm}^{-1}}

\newcommand{\fmiq}{\, \text{fm}^{-3}}

\newcommand{\MeV}{\, \text{MeV}}

\newcommand{\vLambda}{\vb*{\Lambda}}
\newcommand{\vSigma}{\vb*{\Sigma}}
\newcommand{\vPsi}{\vb*{\Psi}}
\newcommand{\vQ}{\vb{Q}}
\newcommand{\vmu}{\vb*{\mu}}
\newcommand{\vx}{\vb{x}}

\newcommand{\vR}{\vb{R}}
\newcommand{\vxbar}{\vb{\bar{x}}}

\newcommand{\rom}[1]{\uppercase\expandafter{\romannumeral #1\relax}}

\newcommand{\YObsi}{\vb{y}_i}

\newcommand{\yi}{\vb{y}_i}
\newcommand{\Yc}{\vb*{\mathcal{Y}}}
\newcommand{\Yci}{\Yc_i}
\newcommand{\vy}{\vb{y}}
\newcommand{\mods}{\mathcal{M}}
\newcommand{\nb}{n_b}
\newcommand{\orcid}[1]{\href{https://orcid.org/#1}{\includegraphics[scale=0.055]{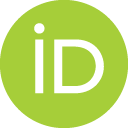}}}

\newcommand{\rev}[1]{#1}  

\begin{document}

\title{A Bayesian mixture model approach to quantifying\\the empirical nuclear saturation point}

\author{C.~Drischler 
\orcid{0000-0003-1534-6285}}
\email{drischler@ohio.edu}
\affiliation{Department of Physics and Astronomy, Ohio University, Athens, OH~45701, USA}
\affiliation{Facility for Rare Isotope Beams, Michigan State University, East Lansing, MI~48824, USA}

\author{P.~G.~Giuliani 
\orcid{0000-0002-8145-0745}}
\email{giulianp@frib.msu.edu}
\affiliation{Facility for Rare Isotope Beams, Michigan State University, East Lansing, MI~48824, USA}
\affiliation{Department of Statistics and Probability, Michigan State University, East Lansing, MI~48824, USA}

\author{S.~Bezoui}
\email{saad.bezoui07@myhunter.cuny.edu}
\affiliation{CUNY Hunter College, 695 Park Ave, New York, NY~10065, USA}

\author{J.~Piekarewicz
\orcid{0000-0003-4980-5670}}
\email{jpiekarewicz@fsu.edu}
\affiliation{Department of Physics, Florida State University, Tallahassee, FL~32306, USA}

\author{F.~Viens 
\orcid{0000-0001-9162-8212}}
\email{viens@rice.edu}
\affiliation{Department of Statistics, Rice University, Houston, TX~77005, USA}

\date{\today}

\begin{abstract}  

The equation of state (EOS) in the limit of infinite symmetric nuclear matter exhibits an equilibrium density, $n_0 \approx 0.16 \, \mathrm{fm}^{-3}$, at which the pressure vanishes and the energy per particle attains its minimum, $E_0 \approx -16 \, \mathrm{MeV}$.  Although not directly measurable, the nuclear saturation point $(n_0,E_0)$ can be extrapolated by density functional theory (DFT), providing tight constraints for microscopic interactions derived from chiral effective field theory (EFT). However, when considering several DFT predictions for $(n_0,E_0)$ from Skyrme and Relativistic Mean Field (RMF) models together, a discrepancy between these model classes emerges at high confidence levels that each model prediction's uncertainty cannot explain. How can we leverage these DFT constraints to rigorously benchmark nuclear saturation properties of chiral interactions? To address this question, we present a Bayesian mixture model that combines multiple DFT predictions for $(n_0,E_0)$ using an efficient conjugate prior approach. The inferred posterior distribution for the saturation point's mean and covariance matrix follows a Normal-inverse-Wishart class, resulting in posterior predictives in the form of correlated, bivariate $t$-distributions.  The DFT uncertainty reports are then used to mix these posteriors using an ordinary Monte Carlo approach. At the 95\% credibility level, we estimate $n_0 \approx 0.157 \pm 0.010 \, \mathrm{fm}^{-3}$ and $E_0 \approx -15.97 \pm 0.40 \, \mathrm{MeV}$ for the marginal (univariate) $t$-distributions. Combined with chiral EFT calculations of the pure neutron matter EOS, we obtain bivariate normal distributions for the nuclear symmetry energy and its slope parameter evaluated at $n_0$: $S_v \approx 32.0 \pm 1.1 \, \mathrm{MeV}$ and $L\approx 52.6\pm 8.1 \, \mathrm{MeV}$ (95\%), respectively. Our Bayesian framework is publicly available, so practitioners can readily use and extend our results.

\end{abstract}

\maketitle

\section{Introduction} 
\label{sec:intro}

Nuclear matter, a strongly interacting system made of protons and neutrons, is pervasive throughout the visible universe.
While the strong interaction binds nucleons into atomic nuclei, it provides the pressure required to support neutron stars against gravitational collapse, driving nuclear matter in their cores to densities far beyond those probed in atomic nuclei. 
Understanding how the properties of nuclear matter emerge and evolve 
across such a wide range of densities has been a critical subject of study, with tremendous progress achieved recently in, \eg, \textit{ab~initio} many-body calculations using microscopic nuclear interactions 
derived from chiral effective field theory (EFT)~\cite{Hammer:2019poc,Tews:2020hgp,Hergert:2020bxy,Hebeler:2020ocj,Drischler:2021kxf,Hu:2021trw,Ekstrom:2022yea,Machleidt:2024bwl}, Bayesian quantification of theoretical uncertainties~\cite{Melendez:2019izc,Drischler:2020yad,Drischler:2020hwi,Drischler:2022ipa,Jiang:2022oba,Jiang:2022tzf,Semposki:2024vnp}, heavy-ion collisions~\cite{Huth:2021bsp,Sorensen:2023zkk,MUSES:2023hyz}, parity-violating electron scattering~\cite{Adhikari:2021phr,CREX:2022kgg,Reed:2021nqk}, and multi-messenger astronomy~\cite{Vitale:2020rid,Corsi:2024vvr,Koehn:2024set}.

A system that has traditionally served to isolate the complex nuclear dynamics is infinite nuclear matter, an idealized framework in which nucleons interact only via the strong force, thereby allowing one to neglect surface effects and thus focus on bulk nuclear matter properties. 
The nuclear equation of state (EOS) encodes how the energy per particle $E(n_b, \delta, T)/A$ changes as a function of the nucleon (number) density $n_b$, isospin asymmetry $\delta$, and temperature $T$. Among the hallmarks of the nuclear dynamics is the saturation property of isospin symmetric nuclear matter (SNM) at zero temperature ($T=0$), characterized by equal neutron and proton densities ($\delta = 0$). In that limit, the EOS exhibits a minimum at the density $n_0\!\approx\!0.16 \fmiq$, which is related to the typical central density of heavy  nuclei~\cite{Horowitz:2020evx}, while the corresponding ground-state energy per particle 
$E(n_0)/A \equiv E_0 \!\approx\!-16 \MeV$ is closely related to the volume term of the semi-empirical 
mass formula. Together, $(n_0,E_0)$ are referred to as the nuclear saturation point. Despite these connections 
to finite nuclei, the saturation point is not a physical observable that could be measured in the laboratory 
(see also Ref.~\cite{Margueron:2017eqc}). However, density functional theory (DFT) extrapolations to infinite 
nuclear matter, informed by nuclear observables, provide important empirical constraints for $(n_0,E_0)$ and 
for other important low-density EOS parameters\rev{~\cite{Bender:2003jk,Chen:2014sca,Reinhard:2005nj,Reinhard:2016mdi,Roca-Maza:2018ujj}}, 
such as the symmetry energy $J\!\equiv\!S_v$ and its slope parameter $L$ evaluated at $n_0$. In turn, 
these bulk parameters guide the construction of microscopic EOS models (\eg, see Refs.~\cite{Hebeler:2013nza,Lim:2018bkq,Huth:2020ozf,Lim:2023dbk}).

Nuclear saturation emerges from a delicate cancellation between kinetic and interaction contributions in the 
microscopic Hamiltonian, which is particularly sensitive to repulsive three-nucleon (3N) interactions in addition 
to the net attractive nucleon-nucleon (NN) contributions (\eg, see Refs.~\cite{Hebeler:2010xb,Hebeler:2020ocj,
Drischler:2021kxf}). This cancellation between two large contributions makes saturation properties 
ideal for benchmarking chiral interactions in the nuclear medium. Moreover, their implementation in 
computational (many-body) frameworks shed light on important open questions in chiral EFT~\cite{Furnstahl:2021rfk}, 
such as Weinberg power counting, the source for the differing predictions for medium-mass to heavy nuclei, 
and the importance of adding the delta-baryon ($\Delta$) degree of freedom. Notably, reproducing the fairly 
tight empirical ranges for $(n_0, E_0)$ obtained from DFT has been challenging for state-of-the-art SNM 
calculations with chiral NN and 3N interactions, even when significant theoretical uncertainties due to EFT 
truncation errors, the use of different chiral interactions, etc. are taken into account (\eg, see Refs.~\cite{Drischler:2015eba,Ekstrom:2015rta,Holt:2016pjb,Drischler:2017wtt,
Lonardoni:2019ypg,Rios:2020oad,Drischler:2021kxf,Sammarruca:2021bpn}). Realistic nuclear saturation 
properties may provide helpful insights into the construction of microscopic nuclear potentials that become more
successful at simultaneously predicting the properties of medium-mass to heavy nuclei and of infinite nuclear
matter~\cite{Simonis:2015vja,Simonis:2017dny,Stroberg:2019bch,Machleidt:2023jws}. 
However, details of the connection between finite nuclei and infinite matter near the saturation point 
have proved elusive~\cite{Hoppe:2019uyw,Huther:2019ont,Atkinson:2020yyo,Hebeler:2020ocj}, so one may need 
further advances in nuclear EFT to properly elucidate them~\cite{Ekstrom:2017koy,Yang:2020pgi,Jiang:2022tzf,Jiang:2022oba,Yang2023,Gasparyan:2023zrf,Thim:2023fnl,Krebs:2023ljo,Krebs:2023gge}.

\begin{figure}[tb]
    \centering
    \includegraphics[width=\linewidth,page=10]{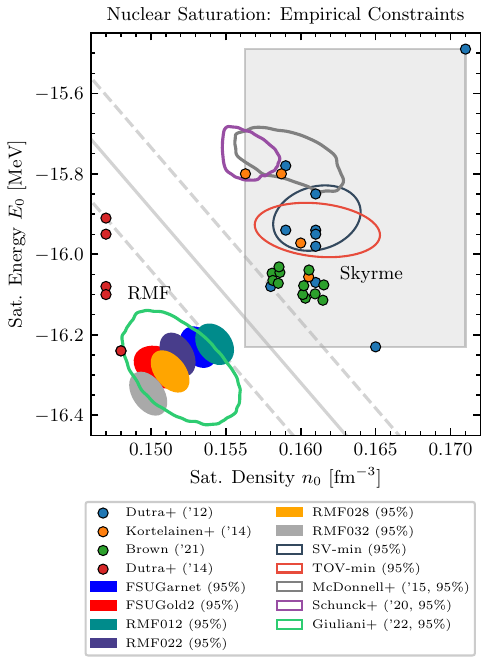}
    \caption{%
   Summary of selected empirical constraints of the nuclear saturation point $(n_0,E_0)$ from Skyrme~\cite{Dutra:2012mb,Kortelainen:2013faa,Klupfel:2008af,Erler:2012qd,Brown:2023private,McDonnell:2015sja,Schunck:2020opr} and relativistic mean field (RMF) models~\cite{Dutra:2014qga,Todd:2003xs,Giuliani:2022yna}. 
    Confidence regions are given at the 95\% credibility level when applicable and unfilled when overlapping.
    The saturation box (gray box) was obtained in Ref.~\cite{Drischler:2015eba} by the range of the 14 data from the sets labeled ``Dutra+''~\cite{Dutra:2012mb} and ``Kortelainen+''~\cite{Kortelainen:2013faa}. The ``+'' in the legend stands for ``\etal''
    This work does not consider the constraints with the solid confidence ellipses in favor of the more recent constraint by Giuliani~\etal~\cite{Giuliani:2022yna}.
    Kernel density estimation (KDE) is used to depict the contours of the constraints by Giuliani~\etal~\cite{Giuliani:2022yna} and Schunck~\etal~\cite{Schunck:2020opr,McDonnell:2015sja}, which are given as random samples from their respective posterior distributions.
    The maximum margin separating hyperplane (MMSH, gray solid line) and its margins (dashed lines), obtained from a linear support vector machine (SVM) trained on all mean values, demonstrate the distinct separation of Skyrme and RMF predictions (see annotations).
    See the main text and Table~\ref{tab:constraints} in Sec.~\ref{sec:results_discussion} for more details.
    } \label{fig:satpoint_constraints}
\end{figure}

Motivated by this narrative, it is natural to ask: \emph{How well constrained is the nuclear saturation point?}
Figure~\ref{fig:satpoint_constraints} summarizes selected empirical constraints of the nuclear saturation point 
$(n_0,E_0)$ from Skyrme~\cite{Dutra:2012mb,Kortelainen:2013faa,Klupfel:2008af,Erler:2012qd,
Brown:2023private,McDonnell:2015sja,Schunck:2020opr} and relativistic mean field (RMF) models~\cite{Dutra:2014qga,Todd:2003xs,Chen:2014sca,Chen:2014mza,Giuliani:2022yna}. 
When available, 95\% confidence regions derived from the developers' uncertainty quantification (UQ) are displayed in the figure.
The gray saturation box depicts the range spanned by the point estimates associated with the Skyrme models compiled in ``Dutra+''~\cite{Dutra:2012mb} and ``Kortelainen+''~\cite{Kortelainen:2013faa} (see also Fig.~\ref{fig:satbox}).
It was constructed in Ref.~\cite{Drischler:2015eba} to serve as a reference for microscopic SNM calculations, 
and a slightly enlarged version was later used to calibrate chiral 3N interactions\,\cite{Drischler:2017wtt}.
Given that all DFT predictions displayed in Fig.~\ref{fig:satpoint_constraints} report saturation 
points that, when considered together, are clearly inconsistent with each other, one must conclude that not all DFT predictions can be both precise \emph{and} accurate simultaneously.

Since, in most cases, the functionals are calibrated to genuine nuclear observables, the source of this inconsistency may be associated with the functional form and/or parameter-estimation protocol of 
the DFT models.\footnote{DFT models may not only be calibrated to finite-nuclei observables. Other constraints 
are informed by neutron star observations and chiral EFT calculations of pure neutron matter; for some recent 
work, see Refs.~\cite{Salinas:2023nci,Alford:2023rgp}).} However, whereas nuclear 
properties such as binding energies and charge radii encode the empirical saturation point, one must recognize 
that an extrapolation to infinite nuclear matter---where Coulomb and surface effects are neglected---may also 
become an additional source of systematic error.  
Although Fig.~\ref{fig:satpoint_constraints} suggests 
relatively small differences in $E_{0}$ because of the large model spread,\footnote{\rev{See also, \eg, Figure~3 in Ref.~\cite{Reinhard:2010ru} for energy trends in super-heavy nuclei based on Skyrme and RMF models.}} there is 
a noticeable difference in the model predictions for the saturation density~\cite{Furnstahl:2001un}. In particular, the Skyrme models 
systematically predict a higher saturation density than the RMF models. The wide margins (dashed 
lines) associated with the maximum margin separating hyperplane (MMSH, gray solid line) of a linear support 
vector machine (SVM) classifier trained on all mean values in Fig.~\ref{fig:satpoint_constraints} demonstrates 
this distinct separation. Additional (unlabeled) DFT constraints with mean values above the MMSH would be 
classified as Skyrme models and those below the MMSH as RMF models (see annotations). 

The difference in the saturation density between the two classes of models is informative. That the 
Skyrme models saturate at higher density suggests that these functionals would, in principle, predict 
smaller charge radii than RMF models. Yet in practice, both sets of models reproduce experimental
charge radii as these are included in the calibration procedure. We must then conclude that to reproduce charge radii, the RMF functionals must have a stiffer surface energy than the Skyrme 
models.
Since infinite nuclear matter is insensitive to surface effects, this may be an important
source behind the discrepancy. In conclusion, there are unavoidable inaccuracies in some, if not all, 
DFT predictions shown in Fig.~\ref{fig:satpoint_constraints} due to underestimated (or even unknown) 
systematic model uncertainties, which are generally difficult to quantify.

\emph{How can we leverage the empirical constraints from DFT to rigorously benchmark nuclear saturation properties of chiral interactions?} To address this question, we propose a Bayesian framework that infers 
the (unobserved) true
empirical saturation point from a collection of pre-selected DFT predictions for 
$(n_0,E_0)$; namely, those represented in Fig.~\ref{fig:satpoint_constraints}. Assuming that all considered 
DFT models provide equally legitimate descriptions of nuclear matter, our framework considers each model's 
prediction for $(n_0, E_0)$ and its stated UQ at face value without further assessment.%
\footnote{\rev{%
 Note that our approach is conceptually and computationally simpler than recent applications of Bayesian model averaging and mixing to nuclear masses (\eg, see Refs.~\cite{Neufcourt:2019sle,Neufcourt:2020nme,Kejzlar:2023tlm}), where the weights with which the to-be-mixed models contribute to the mixed model predictions and additional uncertainty parameters are reevaluated within Bayesian estimation. Here, we assume instead equal weights and do not reevaluate the models' UQ to construct the mixed model.}}
Note that only a few of these DFT constraints include a highly sophisticated method to quantify uncertainties, such as those by Schunck~\etal (Skyrme) via Gaussian process 
(GP) emulation and Giuliani~\etal (RMF) via Galerkin-based reduced order modeling~\cite{Melendez:2022kid,
Bonilla:2022rph,Anderson:2022jhq,Drischler:2022ipa,Duguet:2023wuh}.
The result from our work is a Bayesian mixture model that, given the stated model assumptions and expert knowledge encoded 
in prior distributions, enables statistically robust constraints for the true empirical saturation point, which are 
informed by, and consistent with, the underlying DFT model predictions and their associated uncertainties.
Furthermore, using \rev{\emph{conjugate distributions} associated with the used} likelihood function facilitates a computationally efficient \rev{\emph{conjugate prior approach}} that requires only ordinary Monte Carlo sampling, if uncertainties in the DFT predictions are quantified, and is otherwise analytic
\rev{(see Sec.~\ref{sec:stat_framework} for details)}.
Our framework is publicly available in the form of annotated Jupyter notebooks~\cite{saturationGitHub} 
so that chiral EFT and many-body practitioners can readily use and extend our analysis.

The remainder of this article is organized as follows. Our Bayesian mixture model, its hierarchical design, and 
conjugate distribution approach are presented in Sec.~\ref{sec:stat_framework}. 
The framework is implemented and 
informed by two collections of DFT models in Sec.~\ref{sec:results_discussion}. While the first collection of
models includes only
the Skyrme predictions that are associated with the saturation box in 
Fig.~\ref{fig:satpoint_constraints} (without considering UQ), the second collection includes both Skyrme and RMF models with and 
without UQ. Section~\ref{sec:summary_outlook} concludes with a summary and outlook. Several appendices provide more details.
Appendix~\ref{app:recap_distributions} summarizes key properties of the relevant distribution functions; 
Appendix~\ref{app:conf_region_bivar_t} discusses how confidence regions (\ie, confidence ellipses) of 
bivariate student $t$-distributions can be efficiently computed and plotted; 
Appendix~\ref{app:MC_for_BMM} elaborates on how uncertainty-quantified DFT constraints can be 
incorporated into our conjugate-distribution framework using a Monte Carlo sampling implementation 
of the model mixing approach; and Appendix~\ref{app:suppl_results} provides additional results.
Throughout the article, vectors and matrices are typeset in boldface.

\section{Statistical framework} \label{sec:stat_framework} 

\begin{table*}[tb]
\renewcommand{\arraystretch}{1.2}
\caption{
Notation used in this work.
}
\label{tab:notation}
\begin{ruledtabular}
\begin{tabular}{lp{14.6cm}} 
Notation & Description \\ 
\colrule 
$(n_0,E_0)$ & nuclear saturation point: saturation density $n_0$ (in $\fmiq$) and associated saturation energy (per particle) $E_0 \equiv E(n_0)/A$ (in MeV) of infinite symmetric nuclear matter \\
$d\equiv 2$ & dimensionality of the saturation point\\
$M_i$, $\mathcal{M}$ & collection of one or more DFT constraints, here referred to as model; $\mods = \{M_i\}_{i=1}^n$ \\
$n$ & number of models (\ie, data points) considered in the conjugate prior analysis \\
$\yi\equiv (n_0,E_0)$, $\vb{Y}$ & saturation point of model $M_i$, $i \in [1, 2, \ldots, n]$, modeled as a sample of the random variable $\vb{Y}$ in Eq.~\eqref{eq:hmodel_top_level}; in the statistical meaning typically unobserved, \ie when model $M_i$ provides an uncertainty for $\yi$\\
$\Yci$ & random variable encoding the uncertainties reported with model $M_i$; $\yi$ is treated as a particular sample from $\Yci$; when $M_i$ reports no uncertainty, 
$\Yci=\yi$ is non-random; $\Yc = \{ \Yci\}_{i=1}^n$ \\
$(\vmu, \vSigma)$ & to-be-estimated parameters: mean vector and covariance matrix\footnotemark[1] of the $d$-variate normal distribution $\mathcal{N}_d(\vmu, \vSigma)$ in Eq.~\eqref{eq:hmodel_top_level}\\
$\vb{y}'$ & possible unobserved DFT constraints drawn from the posterior predictive distribution, see Eqs.~\eqref{eq:posterior_pred} and~\eqref{eq:full_posterior_pred}; when $n=0$, these distributions are referred to as prior predictive distributions\\
$t_\nu(\vmu, \vPsi)$ & multivariate $t$-distribution with mean vector $\vmu$ and scale matrix\footnotemark[1]  $\vPsi$; for brevity, this notation is used interchangeably for the name of the distribution and the formula for its density\\
%
%
$\kappa_n, \nu_n, \vmu_n,\vPsi_n$  & hyperparameters of the posterior ($n>0$) or prior ($n=0$), both Normal-inverse-Wishart (NIW) distributions\\
%
%
$X\sim P$   & indicates that the random variable $X$ has the distribution called $P$ (or has the density $P$); \eg, see Eq.~\eqref{eq:hmodel_top_level} where the random variable $\vb{Y}$ has a $d$-variate normal distribution (density)\footnotemark[2] \\
$P(X \mid Y)$ & probability density for the random variable $X$ given (conditional on) another random variable $Y$; this notation can also be used for random vectors
\end{tabular}
\end{ruledtabular}
\footnotetext[1]{The covariance matrix $\vSigma$ and scale matrix $\vPsi$ are real-valued, symmetric, and positive definite.}
\footnotetext[2]{The capital letter $P$ is used to denote probability densities, as is common in the physics literature. In statistics, however, the capitalization is often reserved for actual probabilities, not densities, but this is merely a convention.}
\end{table*}

In this section, we discuss our statistical framework for modeling the empirical saturation point using one or more collections of DFT constraints (see Fig.~\ref{fig:satpoint_constraints}). 
These collections will be referred to as (statistical) models
and will be further specified in Sec.~\ref{sec:results_discussion}.
Table~\ref{tab:notation} summarizes the notation we use throughout this work.

\subsection{Hierarchical Model}
\label{sec:stat_framework_model} 

Our analysis of the empirical saturation point is based on a simple hierarchical model in which
the analysis reported for each functional employed to produce a saturation point is taken at face value, including its reported uncertainty. 
Since we do not have, or seek to obtain, any information on how these various analyses may relate to each other, and since we do not have any external data conferring additional insights about their collective behavior, we assume that each reported saturation point is a given data point, with a given associated uncertainty report.
Therefore, our analysis assumes that the (typically unobserved) saturation point $\yi\equiv (n_0,E_0)$ associated with the model $M_i$ is a sample from a universal $d$-dimensional normal random variable $\vb{Y}$ modeling the true saturation point of all DFT model predictions taken collectively:
\begin{equation} \label{eq:hmodel_top_level}
    \vb{Y} \sim \mathcal{N}_d(\vmu,\vSigma) \,, 
\end{equation}
where $\vmu$ and $\vSigma$ are respectively the \textit{a~priori} unknown mean vector of length $d$ and covariance matrix of size $d\times d$. 
While one has $ d \equiv 2$ for the saturation point, our framework is also applicable to the general $d$-dimensional case. 
Notice that Eq.~\eqref{eq:hmodel_top_level} 
is a classical likelihood model for Bayesian analyses. 
We assume that $\vb{Y}$ is normal for simplicity of analysis, to provide practitioners with as standard a framework as possible, and because the saturation point is a point in a continuous $d$-dimensional space that is only known with some uncertainty. 
Predictions from $n$ different models correspond to repeated draws $\{\yi\}_{i=1}^n$, which are assumed to be independent and identically distributed (i.i.d.).
However, $\{\yi\}_{i=1}^n$ are typically \emph{unobserved} (in the statistical meaning) and thus not directly accessible as data to our analysis due to uncertainties inherent in (and typically reported by) each model in the DFT constraints.

To account for these uncertainties, we introduce instead another $d$-dimensional random variable $\Yci$ whose distribution function $P(\Yci)$ is defined via the model developers' UQ: 
\begin{equation} \label{eq:hmodel_base_level}
    \Yci \sim P(\Yci) \,,~ i=1,2,\ldots ,n \,.
\end{equation} 
As mentioned, each unobserved saturation point $\yi$ is assumed to be a sample coming from the overall model in Eq.~\eqref{eq:hmodel_top_level}, and each individual developer's model $M_i$ produces an uncertain observation distribution $\Yci$ instead of $\yi$.
As illustrated in Fig.~\ref{fig:satpoint_constraints}, the $\Yci$'s considered in this work are:
%
\begin{enumerate}
    %
    \item linear combinations of Dirac $\delta$ distributions for models consisting of one or more point estimates without UQ (\eg, the datasets compiled by Dutra~\etal).
    In the absence of additional information, the point estimates are weighted equally here;
    \item bivariate normal distributions with known mean vectors and covariance matrices (\eg, the functional TOV--min) inferred from covariance analyses; and 
    \item more intricate distributions due to rigorous UQ (\eg, by Giuliani~\etal).
    Since their closed mathematical expressions are unknown or do not exist, numerous samples are used to represent the distributions, which
    are then formally (but not conceptually) equivalent to linear combinations of thousands of Dirac $\delta$ distributions.
\end{enumerate}
%
%
We leave the specific form of these distributions to the expert knowledge of the model developers and take their UQ information, if provided, 
so that all distributions $\vb*{\mathcal{Y}}_i$ are given and need not to be estimated in our Bayesian analysis.
Furthermore, we assume the pairs $(\vb*{\mathcal{Y}}_i,\vb*{\mathcal{Y}}_j)$ to be mutually independent of one another for $i\neq j$, which is somewhat justified since our analysis includes DFT constraints from various sources.
These assumptions are crucial for the streamlined method used in this work. 

\subsection{\rev{Inferring the posterior distribution}}
\label{sec:stat_framework_infer} 

Our task is then to infer the distribution for our model parameters $\vmu,\vSigma$ conditional on the information $\mods$ coming from all the models $\{M_i\}_{i=1}^n$ after averaging over the distributions of all uncertainties $\Yc = \{ \Yci\}_{i=1}^n$:\footnote{%
The notation $P(\Yc = \vy)$ is commonly used in some application areas of statistics to denote the density of the random variable $\Yc$ at the point $\vy$. 
Some statisticians might reserve this notation for discrete random variables, and would use $P_{\Yc}( \vy)$ here instead.}
\begin{equation}
    \begin{split} 
    P(\vmu, & \vSigma \mid \mods) \\
    &= \int \dd{\vy} P(\vmu,\vSigma \mid  \mods, \Yc = \vy) \, P(\Yc = \vy)\,.\label{eq:marginal_post}
\end{split}
\end{equation}
We can interpret Eq.~\eqref{eq:marginal_post}\footnote{We use the notation ``$\mid\mods )$'' in Eq.~\eqref{eq:marginal_post} and throughout the article. 
It is shorthand for the act of conditioning by the information contained in the UQ reported by the developers of DFT model $M_i$ for each $i=1,2,\ldots,n$. 
A more complete notation would require defining probability-measure-valued random elements representing unspecified DFT UQ, and setting those random elements equal to the reported uncertainties. 
Such formalism would be burdensome and entirely unnecessary.} as a posterior distribution, which, using Bayes' theorem,
\begin{equation} \label{eq:bayes_theorem}
    P(\vmu,\vSigma \mid \mods)  = \frac{P(\mods \mid \vmu, \vSigma) \, P(\vmu, \vSigma)}{P(\mods)} \,, 
\end{equation}
ought to be expressed as the product of the likelihood of the information that we have, given the parameter values $\vmu,\vSigma$, times our prior knowledge on these parameters encoded in a distribution $P(\vmu, \vSigma)$ of our choosing, divided by a normalizing constant $P(\mods)$, which is interpreted as an evidence factor in favor of the information $\mods$ at hand. 
Combining Eqs.~\eqref{eq:marginal_post} and~\eqref{eq:bayes_theorem}, one notes that the likelihood of $\mods$ given the parameter values $\vmu,\vSigma$ can be expressed as the following mixing distribution over the model reporting uncertainties: 
\begin{equation}\label{eq:marg_likelihood}
\begin{split}
    P(\mods & \mid \vmu, \vSigma) \\
&= \int \dd{\vy}  P(\mods \mid \vmu, \vSigma, \Yc = \vy) \, P(\Yc = \vy) \,. 
\end{split}
\end{equation}
Note that the likelihood parameters $\vmu$, $\vSigma$ in Eq.~\eqref{eq:bayes_theorem} correspond to our model~\eqref{eq:hmodel_top_level}, and also note that 
\begin{equation} \label{eq:mix_dist}
    P(\Yc) = \prod_{i=1}^n P_i(\Yci) 
\end{equation}
in Eq.~\eqref{eq:marg_likelihood} because of the discussed independence assumption. This interpretation of $\vmu$ and $\vSigma$ implies that our model~\eqref{eq:hmodel_top_level} is indeed our basic likelihood model before acknowledging the uncertainty in the actual sampled saturation points.
To consider this uncertainty, one only needs to mix the basic likelihood against the distributions $P_i(\Yci)$ of the models' observations reported with their UQ. 
This is precisely the operation in Eqs.~\eqref{eq:marginal_post} and~\eqref{eq:marg_likelihood}.  

Alternatively, and with equal validity, we can interpret Eq.~\eqref{eq:marginal_post} as a simple mixture model, where the mixture operation with respect to the mixing variable $\Yc$ (\ie, to its mixing density $P(\Yc=\vy)$) commutes with the application of Bayes' theorem, leading to the following rewriting of that theorem from Eq.~\eqref{eq:bayes_theorem}, in which mixing appears simply as integrating against the density of the random variables $\Yc$: 
\begin{align} 
    P(\vmu, & \vSigma \mid \mods) \notag \\
 &=
 \int \dd{\vy} \,
 P(\vmu, \vSigma \mid \mods, \Yc = \vy ) \, P(\Yc = \vy) \notag \\
 &= 
 \int \dd{\vy} \,
 \frac{ P(\mods \mid \vmu, \vSigma, \Yc) P(\vmu, \vSigma)}{P(\mods)} P(\Yc = \vy)  \,. \label{eq:mixing_post} 
\end{align}
Note that Bayes' theorem, from the second line to the third line in Eq.~\eqref{eq:mixing_post}, is only applied to the parameter-data pair $\left((\vmu, \vSigma), \mods\right)$. 
This insensitivity of the developers' UQ to the Bayesian analysis [\ie, we do not attempt to revise the developers' UQ] is why Bayes' theorem and the mixing operations commute and why Eq.~\eqref{eq:marginal_post} 
is equivalent to Eq.~\eqref{eq:mixing_post}. 
Although formally equivalent, we argue in Sec.~\ref{sec:stat_framework_w_uncert} in favor of using this simple model mixing approach in the last line in Eq.~\eqref{eq:mixing_post} over mixing the likelihood function Eq.~\eqref{eq:marg_likelihood} explicitly to account for uncertainties in the model predictions.

\rev{The choice of the prior distribution $P(\vmu, \vSigma)$ could, in principle, be arbitrary and fully informed by nuclear physics knowledge, leaving significant freedom in the prior choice to the Bayesian practitioner.
Here, we choose a particular prior distribution class that,
for the likelihood function associated with our statistical model~\eqref{eq:hmodel_top_level}, ensures that the prior and posterior distributions are in the same distribution class.
With respect to that likelihood function, the prior distribution is then referred to as the \emph{conjugate prior} and the prior and posterior distributions together as \emph{conjugate distributions}.\footnote{\rev{See, e.g., Section~2.4 in Ref.~\cite{gelman2013bayesian}, Section~4.6.1 in Ref.~\cite{pml1Book}, and Definition~4 in Ref.~\cite{hoff2009first} for more details on conjugate priors and conjugate distributions.}} 
This classical \emph{conjugate prior approach}~\cite{pml1Book,gelman2013bayesian,hoff2009first} has the computationally desirable and convenient feature that the posterior parameters have closed-form expressions that can be inferred analytically via simple algebra,  without needing (computationally more demanding) Monte Carlo sampling or other computational approximations that would be required if nonconjugate distributions were used. 
As detailed in the following Sec.~\ref{sec:stat_framework_wo_uncert}, in our case, the conjugate distributions with respect to our normal likelihood function~\eqref{eq:likelihood} are Normal-inverse Wishart (NIW) distributions.
}


\subsection{DFT constraints without uncertainties}
\label{sec:stat_framework_wo_uncert} 

We discuss first the case in which $\{\yi \}_{i=1}^n$ is \emph{observed}, \ie, each model contains only a point estimate for $\yi$ without (quantified) uncertainties.
In this scenario, we can symbolically evaluate the mixture model~\eqref{eq:marginal_post}, or likewise the marginalized likelihood~\eqref{eq:marg_likelihood}, since Eq.~\eqref{eq:mix_dist} simplifies to a product of Dirac $\delta$ distributions located at the (in this case) observed true saturation point $\YObsi$,
\begin{equation}
    P(\Yc) = \prod_{i=1}^n \delta (\Yci - \YObsi) \,.
\end{equation}
We encounter this scenario in Sec.~\ref{sec:results_satbox_analysis} when we analyze the datasets used in Ref.~\cite{Drischler:2015eba} to construct the saturation box in Fig.~\ref{fig:satpoint_constraints}.
The likelihood function for the observed data, corresponding to our statistical model~\eqref{eq:hmodel_top_level}, is then given by the multivariate normal distribution
\begin{multline} \label{eq:likelihood}
        P(\mods \mid \vmu, \vSigma)  = \prod_{n=1}^n \mathcal{N}_d(\YObsi \mid \vmu, \vSigma) \\ 
        \propto |\vSigma|^{-\frac{n}{2}} \exp \left[-\frac{1}{2}\sum \limits_{i=1}^n  (\YObsi- \boldsymbol{\mu})^{\intercal} \vSigma^{-1}(\YObsi- \boldsymbol{\mu}) \right] \,,
\end{multline}
with the determinant functional $|\bullet|$ and number of data points $n$ considered. 
Note that the product, the sum inside the exponential function, and the term $|\vSigma|^{-\frac{n}{2}}$ in Eq.~\eqref{eq:likelihood}, reflect the assumption that the samples $\YObsi$ are i.i.d. draws of the random variable~\eqref{eq:hmodel_top_level}. 
\rev{To take advantage of the computational efficiency of conjugate distributions, we use here the conjugate prior with respect to the normal likelihood function~\eqref{eq:likelihood}: the Normal-inverse-Wishart (NIW) distribution~\cite{gelman2013bayesian,pml1Book,pml2Book}}  
%
\begin{align}
    P(\vmu, \vSigma) &= \operatorname{NIW}(\vmu,\vSigma \mid  \kappa_0,\nu_0,\vmu_0,\vPsi_0) \notag \\
    &\propto
    |\vSigma|^{-\left(\left(\nu_{0}+d\right) / 2+1\right)} \nonumber \\
    &\quad \times \exp \left[-\frac{\kappa_{0}}{2}\widetilde{\vmu}^{\intercal}\vSigma^{-1}\widetilde{\vmu}-\frac{1}{2}\operatorname{tr}(\vPsi_0\vSigma^{-1})\right] \,, \label{eq:prior}
\end{align}
with the shorthand notation $\widetilde{\vmu} = (\vmu-\vmu_0)$ and trace operator $\tr(\bullet)$ on the space of $d\times d$ matrices.
The NIW distribution~\eqref{eq:prior} is parametrized in terms of the four hyperparameters $\left(\kappa_{0}, \nu_{0}, \vmu_{0}, \vPsi_{0} \right)$, where the integers $\kappa_{0}>0$ and $\nu_{0}>0$ respectively are a scale parameter and the degree of freedom, while $\vmu_0$ is the prior mean vector and $\vPsi_{0}$ the (real-valued, symmetric, positive-definite) prior scale matrix.
Smaller values for $(\kappa_0, \nu_0)$ result in heavier tails at the prior level and thus typically also in more conservative UQ, particularly in data-limited settings like the one considered here. 
A slightly more conservative UQ than required can be a virtue to guard against
under-reporting uncertainty.

\rev{Using the normal likelihood function~\eqref{eq:likelihood} and the NIW prior~\eqref{eq:prior}, the posterior distribution is also NIW-distributed and given by~\cite{gelman2013bayesian,pml1Book,pml2Book}}
\begin{equation} \label{eq:posterior}
    P(\vmu, \vSigma \mid \mods) =     \operatorname{NIW}(\vmu,\vSigma \mid  \kappa_n,\nu_n,\vmu_n,\vPsi_n; \mods) \,,
\end{equation}
with the updated hyperparameters
\begin{subequations} \label{eq:updated_params_niw}
\begin{align}
\kappa_{n} &=\kappa_{0}+n\,, \\
\nu_{n} &= \nu_{0}+n\,, \\
\vmu_{n} &= \frac{1}{\kappa_{n}} \big[\kappa_{0}\vmu_0+ n\bar{\vb{y}} \big]\,, \\
\vPsi_{n} &= \vPsi_{0}+\vb{S}+\frac{\kappa_{0} n}{\kappa_{n}} \left(\bar{\vb{y}} - \vmu_0\right) \left(\bar{\vb{y}} - \vmu_0 \right)^{\intercal} \,,
\end{align}
\end{subequations}
where the grand sample mean is denoted by
\begin{equation} \label{eq:grand_sample_mean}
    \bar{\vb{y}} = \frac{1}{n} \sum \limits_{i=1}^n \vb{y}_i\,,
\end{equation}
and sum-of-squared-deviations matrix from the grand sample mean~\eqref{eq:grand_sample_mean} is denoted by\footnote{Note that the $d\times d$ sum-of-squared-deviations matrix~\eqref{eq:S-matrix} divided by $n-1$ corresponds to the (unbiased) sample covariance matrix.}
\begin{equation} \label{eq:S-matrix}
\vb{S}=\sum_{i=1}^{n}\left(\vb{y}_i-\bar{\vb{y}}\right)\left(\vb{y}_i-\bar{\vb{y}}\right)^{\intercal} \,.
\end{equation}
By inspecting Eqs.~\eqref{eq:updated_params_niw}, one can gain some intuition for the meaning of the (positive-integer) hyperparameters $(\kappa_0, \nu_0)$: 
the mean vector $\vmu_n$ is estimated from $\kappa_0$ prior observations with sample mean vector $\vmu_0$, whereas the scale matrix $\vPsi_{n}$ is estimated from $\nu_0$ prior observations with mean vector $\vmu_0$ and scale matrix $\vPsi_0 = \nu_0 \vSigma_0$, where $\vSigma_0$ is interpreted as a prior covariance matrix.

By combining the inferred posterior~\eqref{eq:posterior} with our model assumption~\eqref{eq:hmodel_top_level},
we can determine the distribution of possible future DFT predictions based on the observed model constraints,
\begin{equation} \label{eq:posterior_pred_formal}
P(\vb{y}' \mid \mods) = \int \dd{\vmu} \int \dd{\vSigma} \, \mathcal{N}_d(\vb{y}' \mid \vmu,\vSigma) \, P(\vmu, \vSigma \mid \mods) \,,
\end{equation}
also known as the \emph{posterior predictive distribution} (PPD) of the model's predicted response variable $\vb{y}'$. 
It accounts for the uncertainties in the model parameters $\vmu, \vSigma$ through integrating over the posterior~\eqref{eq:posterior}.
Equation~\eqref{eq:posterior_pred_formal} is another example of a mixture model, where a distribution is obtained from a given probability model, assuming its parameters have their own randomness, independent of the original model. 
In our case, the integral in Eq.~\eqref{eq:posterior_pred_formal} can be carried out analytically (\eg, see Refs.~\cite{gelman2013bayesian,pml2Book}), resulting in a multivariate $t$-distribution $t_{\nu}(\vmu, \vPsi)$ with $\nu=\nu_n-d+1$ degrees of freedom, mean vector $\vmu=\vmu_n$, and scale matrix $\vPsi$ (proportional to the distribution's covariance matrix) given below as a function of the updated hyperparameters in Eqs.~\eqref{eq:updated_params_niw}:
\begin{align}
    P(\vb{y}' \mid \mods) 
    &= t_{\nu_n -d +1} \left( \vmu_n, \frac{\vPsi_n(k_n+1)}{k_n(\nu_n-d+1)} \right) \,. \label{eq:posterior_pred}
\end{align}
In the case $n=0$, we refer to Eq.~\eqref{eq:posterior_pred} as the prior predictive distribution, which is completely data-agnostic but is still interpreted as a prediction from the model using a mixture (\ie, the priors are the mixing distributions).
The marginal distributions of the posterior (and prior) predictive distribution~\eqref{eq:posterior_pred} 
are also multivariate $t$-distributions with the same degree of freedom. 
Specifically, in the bivariate case, the two marginal distributions are the univariate student $t$-distributions $t_\nu(\mu  = \vmu_i,\sigma_i^2=\vPsi_{ii})$ when the saturation density ($i=2$) or the energy ($i=1$) is marginalized out. 
Appendix~\ref{app:recap_distributions} reviews selected properties of the NIW~\eqref{eq:posterior} and multivariate $t$-distribution~\eqref{eq:posterior_pred} and discusses how random samples from these distributions can be generated, \eg, using the Python package \texttt{scipy}~\cite{2020SciPy-NMeth}.
Furthermore, Appendix~\ref{app:conf_region_bivar_t} presents an analytic method to determine confidence regions for bivariate $t$-distributions at given credibility levels and how they can be plotted.

\subsection{DFT constraints with uncertainties}
\label{sec:stat_framework_w_uncert} 

We now revisit the general case in which model uncertainties are quantified; \ie, $P(\Yc_i=\YObsi) \neq \delta(\Yci - \YObsi)$ for at least one model. 
In this scenario, the likelihood function~\eqref{eq:marg_likelihood} has to account for the random vector $\Yc$ explicitly. 
One may consider doing so by calculating our statistical model's full likelihood function~\eqref{eq:marg_likelihood}.
But this approach would require Markov chain Monte Carlo (MCMC) sampling as it breaks the conjugacy needed for the conjugate prior approach discussed in Sec.~\ref{sec:stat_framework_wo_uncert}.

Although brute-force MCMC sampling of the posterior~\eqref{eq:bayes_theorem} is feasible, we argue here in favor of the alternative and computationally less expensive approach in which we interpret the desired posterior~\eqref{eq:marginal_post} as a mixture model with respect to the random variable $\Yc$, as discussed in Sec.~\ref{sec:stat_framework_model}.
Because the to-be-mixed posterior~\eqref{eq:mixing_post} is conditional on $\Yc$, we can apply the conjugate prior approach in Sec.~\ref{sec:stat_framework_wo_uncert} to obtain
\begin{equation} \label{eq:full_posterior}
    P(\vmu, \vSigma \mid \mods, \Yc) \\ =     
    \operatorname{NIW}(\vmu,\vSigma \mid  \kappa_n,\nu_n,\vmu_n,\vPsi_n; \mods, \Yc) \,,
\end{equation}
with the updated posterior hyperparameters given by Eqs.~\eqref{eq:updated_params_niw} and $\Yc=\{\YObsi\}_{i=1}^n$, where each $\YObsi$ is a sample from the distribution of the random variable  $\Yci$ provided by the developers of model $\mods_i$.
The desired posterior $P(\vmu, \vSigma \mid \mods)$ is then obtained by integrating against the distribution of the mixing model~\eqref{eq:marginal_post} in the conditional posterior Eq.~\eqref{eq:full_posterior}:
\begin{multline} \label{eq:full_posterior_unconditional}
P(\vmu, \vSigma \mid \mods) = \int \dd{\vy} P(\Yc=\vy) \\
     \times \operatorname{NIW}(\vmu,\vSigma \mid  \kappa_n,\nu_n,\vmu_n,\vPsi_n; \mods, \Yc=\vy) \,. 
\end{multline}
Furthermore, model mixing at the level of the posterior propagates to the posterior (and prior) predictive distribution:
\begin{multline} \label{eq:full_posterior_pred}
    P(\vb{y}' \mid \mods)
    = \int \dd{\vy} P(\Yc=\vy) \\
     \quad \times t_{\nu_n -d +1} \left( \vmu_n, \frac{\vPsi_n(k_n+1)}{k_n(\nu_n-d+1)} \mid \mods, \Yc = \vy\right)  \,. 
\end{multline}
Note that Eq.~\eqref{eq:full_posterior_pred} simplifies to Eq.~\eqref{eq:posterior_pred} in the case of vanishing uncertainties discussed in Sec.~\ref{sec:stat_framework_wo_uncert}.
Integrating against $\dd{\vy} P(\Yc=\vy)$ in Eqs.~\eqref{eq:full_posterior_unconditional} and~\eqref{eq:full_posterior_pred} (and also formally in Eq.~\eqref{eq:mixing_post}) can also be interpreted as computing the mathematical expectation (\ie, the expected value) with respect to the randomness in $\Yc$, which is a consequence of the fact that our hierarchical model is a mixture. 

We use here the following iterative process to sample from the posterior predictive distribution derived in Eq.~\eqref{eq:full_posterior_pred} for our Bayesian analysis. 
First, we sample each $\Yci$, with $i \in [1, 2, \ldots, n]$, once from their respective probability distributions.
Treating these samples from $\Yc$ as observed data $\YObsi$, we can apply the conjugate prior approach discussed in Sec.~\ref{sec:stat_framework_wo_uncert} to obtain a sample from the posterior predictive distribution~\eqref{eq:posterior_pred} when all the $\YObsi$'s are known.
Then, we draw $M>0$ samples $\vb{y}'$ from this posterior predictive distribution~\eqref{eq:posterior_pred}. 
We repeat this process $N>0$ times, including the re-sampling of $\YObsi$ in each of the $N-1$ additional iterations, until the desired total number of sampling points (of the posterior predictive) $Q := M\times N$ is obtained.

We choose $Q=10^8$ to explore the uncertainties in the statistical models and $N=10^{6}$ to explore the DFT developers' UQ. 
This corresponds to $M=10^{2}$. 
A large value for the total number of samples $Q$ is particularly judicious because the posterior predictions are $t$-distributed, with moderate values for their posterior degree of freedom, implying that the distribution has relatively heavy tails. 
This tail property, which reflects the data-limited aspect of this study, can only be well-explored with a sufficiently large number of samples. 
We have checked that using $Q=10^{8}$ sampling points is a very conservative choice.
The comparatively modest number $N$ of samples used to explore the various models' DFT prediction uncertainties is a parsimonious choice, reflecting the fact that most of the procedure's overall uncertainty comes from the discrepancy between the various model predictions, with the individual model uncertainty appearing as a secondary source in the overall UQ. 

This iterative process is an instance of an ordinary Monte Carlo method, with no dependence across iterations, unlike MCMC methods, making this process embarrassingly parallelizable~\cite{Herlihy2021iv}. 
Mathematical details on this ordinary Monte Carlo approach with variance reduction can be found in Appendix~\ref{app:MC_for_BMM}\rev{; see also Chapter~5 in Ref.~\cite{VarRedChapter5}}.
Other choices for sampling from Eq.~\eqref{eq:full_posterior_pred} exist, including using a single iterative loop where only a single new sample (\ie, $M=1$) is used for each $\YObsi$ at each iteration. 
The choice $M=1$ can be described as the direct instance of a Monte Carlo approach for model mixing. 
However, we have found that our choice of $(M,N)$ is most computationally efficient \rev{in obtaining} a stable posterior prediction without needing an inordinate amount of samples.
\section{Results and Discussion} 
\label{sec:results_discussion}

\begin{table*}[tb]
\renewcommand{\arraystretch}{1.2}
\caption{%
Details of the DFT constraints shown in Fig.~\ref{fig:satpoint_constraints}, including Skyrme and RMF models.
The constraints are given as a set of a few individual data points (\ie, ``Dutra+~('12))'', as bivariate normal distributions (\ie, ``TOV--min''), or as a large collection of data points obtained from posterior sampling (\ie, ``Schunck+~('20)'').
The number of constraints (or random samples) is denoted by $n_i$, and whether the DFT models are uncertainty quantified is specified in the column ``UQ.''
}
\label{tab:constraints}
\begin{ruledtabular}
\begin{tabular}{lllllp{6.9cm}}
Label & Type & Given as & $n_i$ & UQ? & Comments\\ 
\colrule
Dutra+ ('12)~\cite{Dutra:2012mb} & Skyrme & empirical & 10 & no & this comprehensive analysis selected 16 out of 240 Skyrme models; see Table~VII in Ref.~\cite{Dutra:2012mb}; six of them were excluded due to their unreasonable behavior at high densities or being unstable for finite nuclei~\cite{Dutra:2012mb,Brown:2013pwa}; the remaining models are: 
KDE0v1, NRAPR, Ska25, Ska35, SKRA, SkT1, SkT2, SkT3, SQMC700, and SV--sym32 \\
Kortelainen+ ('14)~\cite{Kortelainen:2013faa} & Skyrme & empirical & 4 & yes\footnotemark[1] & the models are: SLy4, UNEDF0, UNEDF1, and UNEDF2; see \mbox{Table~IV} in Ref.~\cite{Kortelainen:2013faa}\\
Brown ('21)~\cite{Brown:2023private} & Skyrme & empirical & 12 & no & informed by microscopic calculations of low-density neutron matter; see Ref.~\cite{Brown:2013pwa} for details \\
Dutra+ ('14)~\cite{Dutra:2014qga} & RMF & empirical & 5 & no & this comprehensive analysis of 263 RMF models (similar to Ref.~\cite{Dutra:2012mb}) selected BKA22, BKA24, BSR11, BSR12, Z271v5, and Z271v6; see Table~VI in Ref.\,\cite{Dutra:2014qga} \\ 
FSUGarnet~\cite{Chen:2014mza}\footnotemark[2] & RMF & Normal & 1 & yes & calibrated following the same procedure as Ref.~\cite{Chen:2014sca}, but with the additional constraint of reproducing the existence of recently observed neutron-rich isotopes. Covariance analysis between model parameters and nuclear properties are provided within a Gaussian approximation \\
FSUGold2~\cite{Chen:2014sca}\footnotemark[2] & RMF & Normal & 1 & yes & calibrated using nuclear masses, radii, monopole responses, and neutron star properties. Covariance analysis between model parameters and nuclear properties is provided within a Gaussian approximation.\\
RMF0XY~\cite{Chen:2014mza}\footnotemark[2]\footnotemark[3] & RMF & Normal & 5 & yes & calibrated following the same procedure as Ref.~\cite{Chen:2014mza}, but with the additional constraints of different assumed values for the neutron skin of $^{208}$Pb. Covariance analysis between model parameters and nuclear properties is provided within a Gaussian approximation \\
SV--min~\cite{Klupfel:2008af} & Skyrme & Normal & 1 & yes &  calibration informed by ground-state properties, including binding energies and radii, of various nuclei  \\
TOV--min~\cite{Erler:2012qd} & Skyrme & Normal & 1 & yes & calibration additionally constrained by neutron stars; similar to SV--min but with different isovector properties
\\
McDonnell+ ('15)~\cite{McDonnell:2015sja} & Skyrme & empirical\footnotemark[4] & $10^3$ & yes & 
Bayesian calibration of the model in Ref.~\cite{kortelainen2012nuclear} (UNEDF1) using nuclear masses, radii, and fission observables of spherical and deformed nuclei. The posterior is efficiently explored by using a GP emulator~\cite{higdon2008computer}\\
Schunck+ ('20)~\cite{Schunck:2020opr} & Skyrme & empirical\footnotemark[4] & $6\times10^3$ & yes & 
Bayesian calibration of the model in Ref.~\cite{kortelainen2012nuclear} (UNEDF1) using newly available nuclear masses, radii, and fission observables of spherical and deformed nuclei. The posterior is efficiently explored using a GP emulator~\cite{higdon2008computer}\\
Giuliani+ ('22)~\cite{Giuliani:2022yna} & RMF & empirical\footnotemark[4] & $10^5$ & yes & Bayesian calibration of the model in Ref.~\cite{Todd:2003xs} using nuclear masses and charge radii. The posterior is efficiently explored via sampling using a reduced basis method emulator~\cite{Bonilla:2022rph, Melendez:2022kid}.
\end{tabular}
\footnotetext[1]{For uncertainty estimates, see, \eg, Table~II in Ref.~\cite{kortelainen2012nuclear} (UNEDF1) and Table~III in Ref.~\cite{Kortelainen:2013faa} (UNEDF2). In this work, however, we use the point estimates in Table~IV in Ref.~\cite{Kortelainen:2013faa} to be consistent with Ref.~\cite{Drischler:2015eba}.}
\footnotetext[2]{These constraints are \emph{not} considered here in favor of the more recent constraint by Giuliani~\etal~\cite{Giuliani:2022yna}.
}
\footnotetext[3]{The placeholder ``0XY' stands for a string of three integers, encoding the predicted \isotope[208]{Pb} neutron skin thickness; \ie, $R_\text{skin}^{208} = 0.\text{XY} \, \text{fm}$.}
\footnotetext[4]{These constraints are given as random samples from the authors' posterior distributions.}
\end{ruledtabular}
\end{table*}

In this section, we present the results of our Bayesian inference of the empirical saturation point, which we obtain by applying the Bayesian mixture model discussed in Sec.~\ref{sec:stat_framework} to the various DFT constraints depicted in Fig.~\ref{fig:satpoint_constraints} and detailed in Table~\ref{tab:constraints}.
We also study the implications of our results, combined with recent microscopic pure neutron matter (PNM) calculations, on the nuclear symmetry energy and its density dependence evaluated at the (empirical) saturation density.
Our publicly available Jupyter notebooks~\cite{saturationGitHub} enable the interested reader to experiment with different prior hyperparameters and extend our analysis in various directions.

\subsection{DFT constraints without uncertainties: saturation box}
\label{sec:results_satbox_analysis}

\begin{figure}[tb]
    \centering
    \includegraphics[width=\linewidth]{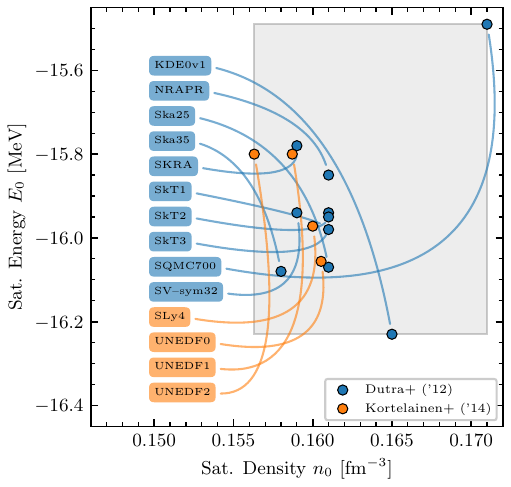}
    \caption{%
     Collection of data points ($n=14$) used in Ref.~\cite{Drischler:2015eba} to construct the saturation box depicted in Fig.~\ref{fig:satpoint_constraints} (gray box).
     It combines the models labeled Dutra~\etal ('12) and Kortelainen~\etal ('14) in Table~\ref{tab:constraints} and thus contains only Skyrme models without considering UQ.
     The curved guidelines connect the labels of the models to their respective data points.
    } \label{fig:satbox}
\end{figure}

We begin our analysis with the collection of DFT models used in Ref.~\cite{Drischler:2015eba} to construct the saturation box, which is depicted in Figs.~\ref{fig:satpoint_constraints} and~\ref{fig:satbox} (see also Section~IV.B in Ref.~\cite{Drischler:2015eba}). 
This box represents the (uncorrelated) range in $(n_0, E_0)$ obtained by the $n=14$ Skyrme models labeled ``Dutra~\etal ('12)'' (blue points) and ``Kortelainen~\etal ('14)'' (orange points) in Table~\ref{tab:constraints}:
\begin{subequations} \label{eq:satbox}
\begin{align}
    n_0^{\text{(box)}} &\approx 0.164 \pm 0.007 \fmiq ,\\ 
    E^{\text{(box)}}_0 &\approx -15.9 \pm 0.4 \MeV \,.
\end{align}
\end{subequations}
Note that these ranges are not associated with any credibility interval, and none of the underlying data points comes with UQ.
They are treated as independent model predictions.
As shown in Fig.~\ref{fig:satbox}, the energy functionals UNEDF2 and KDE0v1 set the lower bound on $n_0$ and $E_0$, respectively, while SQMC700 sets the upper bound on both.
Hence, the saturation box would shrink considerably if SQMC700 were not considered.
\rev{In Appendix~\ref{app:suppl_results}, we provide additional results for our saturation point analysis without SQMC700 and with additional point-like DFT constraints without UQ, including Gogny interactions and a Fayans functional.}

\begin{figure*}[tb]
    \centering
    \includegraphics[width=0.49\linewidth,page=1]{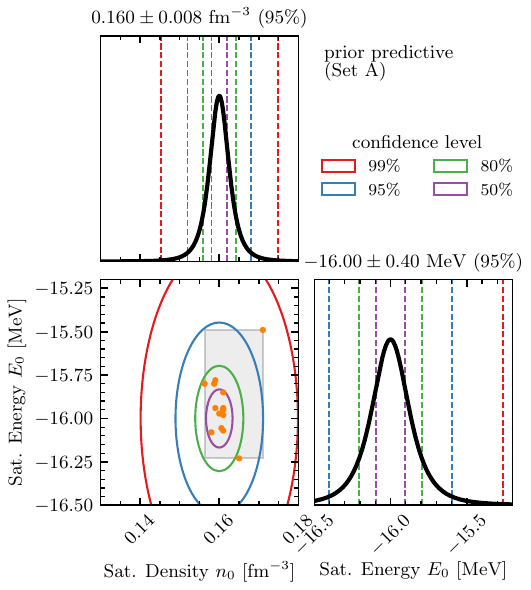}
    \includegraphics[width=0.49\linewidth,page=2]{figures/corner_satbox.pdf}
    \caption{%
    Prior predictive (left panel; data-agnostic) and posterior predictive (right panel; data-informed) and their respective marginal distributions in the panels along the diagonals based on the prior parameters in Set~A.
    The $n=14$ data points (orange points) used to construct the saturation box (see  Fig.~\ref{fig:satbox} for details) are considered; \ie, the datasets labeled Dutra~\etal ('12) and Kortelainen~\etal ('14) in Fig.~\ref{fig:satpoint_constraints} and Table~\ref{tab:constraints}.
    The colored ellipses and dashed vertical lines encompass confidence regions centered at the distributions' mean values at different credibility levels, as specified in the legends. 
    The titles of the panels along the diagonals give the 95\% confidence regions of the marginal distributions, corresponding to the blue vertical lines.
    } \label{fig:satbox_analysis_set_A}
\end{figure*}

With the data collection specified, we need to encode our prior knowledge of the empirical saturation point in the four parameters entering the prior distribution~\eqref{eq:prior}.
We compare the results of two prior parameter sets, labeled Set~A and Set~B, as a simple proxy for the prior sensibility of our results.
The two prior sets use the smallest values possible for the hyperparameters $(\kappa_0=1, \nu_0=4)$ that still support a well-defined expectation value of the underlying inverse-Wishart prior distribution (see Appendix~\ref{app:recap_distributions} for details).
Furthermore, we assume no prior knowledge of the correlation between $(n_0, E_0)$, rendering the scale matrices $\vPsi_0$ in both prior sets diagonal.

For Set~A, we choose the NIW prior mean vector\footnote{This choice of prior mean is informed by the physical reality emerging from what an informal scientific consensus would look like, \eg, in view of Fig.~\ref{fig:satpoint_constraints}. Contrary to a standard Bayesian regression analysis where it is common to choose uninformed prior means equal to zero for all regression coefficients, our model~\eqref{eq:hmodel_top_level} has no regression coefficients, only a single mean vector, making the use of an informed prior mean important.} and scale matrix respectively to be
\begin{equation} \label{eq:prior_setA}
    \vmu_0^{\text{(A)}} \equiv 
    \begin{bmatrix}
    n_0\\ E_0
    \end{bmatrix} =
    \begin{bmatrix}
        0.160 \\ -16.00
    \end{bmatrix} \;, 
    \vPsi_0^{\text{(A)}} = \mqty[\dmat[0]{0.003^2,0.15^2}] \,,
\end{equation}
resulting in the NIW posterior~\eqref{eq:posterior} with the updated hyperparameters $(\kappa_n = 15,\nu_n = 18)$, and
\begin{equation} \label{eq:posterior_setA}
    \vmu_n^{\text{(A)}} \approx \begin{bmatrix}
    0.161 \\ -15.93
    \end{bmatrix} \,, \quad
    \vPsi_n^{\text{(A)}} \approx \mqty[0.013^2 & 0.051^2 \\ 0.051^2 & 0.66^2] \,.
\end{equation}
For brevity, throughout this article, we omit the units for $n_0$~(in $\fmiq$) and $E_0$~(in MeV) when given in matrix-vector notation. 
The parameter choice in Eq.~\eqref{eq:prior_setA} incorporates our prior knowledge that $(n_0,E_0) \approx (0.16 \fmiq, -16 \MeV)$ and that the data point associated with SQMC700 (see Fig.~\ref{fig:satbox}) might be an outlier~\cite{Brown:2023private}.

Figure~\ref{fig:satbox_analysis_set_A} shows the corresponding prior predictive~\eqref{eq:posterior_pred} [\ie, $n=0$] with the prior parameters~\eqref{eq:prior_setA} in the left panel and the posterior predictive~\eqref{eq:posterior_pred} [\ie, $n=14$] with the updated parameters~\eqref{eq:posterior_setA} in the right panel. 
The marginal distributions are plotted along the diagonals of each panel.
Recall that the two predictive distributions are bivariate $t$-distributions, whose marginal distributions are univariate $t$-distributions, as discussed in Sec.~\ref{sec:stat_framework} and Appendix~\ref{app:multvar_t}. 
Although both panels depict the dataset underlying our analysis by the orange points (serving as a helpful reference), we stress that the prior predictive (left panel) is agnostic of these data points.
The data-informed posterior predictive~\eqref{eq:posterior_pred} (right panel) is given by the $t_\nu(\vmu, \vPsi)$ with hyperparameters:
\begin{equation} \label{eq:ppd_setA}
\nu^{\text{(A)}} = 17 \,, \quad 
    \vmu^{\text{(A)}} = \vmu_n^{\text{(A)}} \,, \quad
    \vPsi^{\text{(A)}} \approx \mqty[0.003^2 & 0.013^2 \\ 0.013^2 & 0.17^2] \,.
\end{equation}
To illustrate the tails of the predictive distributions, Fig.~\ref{fig:satbox_analysis_set_A} shows (elliptical) confidence regions centered at the distribution's mean vector,\footnote{The elliptical contour lines at given credibility levels correspond to constant Mahalanobis distances associated with the bivariate $t$-distribution, as discussed in Appendix~\ref{app:conf_region_bivar_t}.} corresponding to the $\{50, 80, 95, 99\}\%$ credibility levels, which are common levels used in UQ studies (see the legends).
Our GitHub repository~\cite{saturationGitHub} provides the Python function \texttt{plot\_confregion\_bivariate\_t($\ldots$)} for calculating and plotting these confidence ellipses efficiently based on the analytic derivation in Appendix~\ref{app:conf_region_bivar_t}.
The dashed vertical lines in Fig.~\ref{fig:satbox_analysis_set_A} highlight the corresponding percentiles of the marginal univariate $t$-distributions, with the associated (blue) 95\% credibility region quoted in the titles of these panels.
That is, for the posterior predictive, we obtain the following marginal 95\% credibility regions:
\begin{subequations} \label{eq:marginal_cred_range_setA}
\begin{align}
    n_0^{\text{(A)}} &\approx 0.161 \pm 0.007 \fmiq \,,\\ 
    E_0^{\text{(A)}} &\approx -15.93 \pm 0.35 \MeV \,.
\end{align}
\end{subequations}
%

For Set~B, we choose the NIW hyperparameters $(\kappa_0 = 1$, $\nu_0 = 4)$ as in Set~A, but design the mean vector and (diagonal) scale matrix to be more weakly informed,
\begin{equation} \label{eq:prior_setB}
    \vmu_0^{\text{(B)}} = \begin{bmatrix}
        0.163 \\ -15.90
    \end{bmatrix} \,, \quad
    \vPsi_0^{\text{(B)}} = \mqty[\dmat[0]{0.004^2,0.28^2}] \,,
\end{equation}
resulting in the NIW posterior~\eqref{eq:posterior} with $(\kappa_n = 15, \nu_n = 18)$, as in Set~A, and
\begin{equation} \label{eq:posterior_setB}
    \vmu_n^{\text{(B)}} \approx \begin{bmatrix}
    0.161 \\ -15.92
    \end{bmatrix} \,, \quad
    \vPsi_n^{\text{(B)}} \approx \mqty[0.013^2 & 0.051^2 \\ 0.051^2 & 0.70^2] \,.
\end{equation}
Figure~\ref{fig:satbox_analysis_set_B} in Appendix~\ref{app:suppl_results} shows the prior predictive and posterior predictive for Set~B similarly to Fig.~\ref{fig:satbox_analysis_set_A} for Set~A. 
The data-informed posterior predictive~\eqref{eq:posterior_pred} (right panel) is given by the $t_\nu(\vmu, \vPsi)$ with hyperparameters:
\begin{equation} \label{eq:ppd_setB}
\nu^{\text{(B)}} = 17 \,, \quad 
    \vmu^{\text{(B)}} = \vmu_n^{\text{(B)}} \,, \quad
    \vPsi^{\text{(B)}} \approx \mqty[0.003^2 & 0.013^2 \\ 0.013^2 & 0.18^2] \,,
\end{equation}
corresponding to the following 95\% confidence regions of the marginal distributions,
\begin{subequations} \label{eq:marginal_cred_range_setB}
\begin{align}
    n_0^{\text{(B)}} &\approx 0.161 \pm 0.007 \fmiq \,,\\ 
    E_0^{\text{(B)}} &\approx -15.92 \pm 0.37 \MeV \,,
\end{align}
\end{subequations}
which are consistent with (but slightly more uncertain in $E_0$ than) those obtained for Set~A in Eq.~\eqref{eq:marginal_cred_range_setA}, as expected from the more weakly informed prior.

Overall, we find for both prior sets that the resulting posterior predictive~\eqref{eq:posterior_pred} is \emph{not} normal (\ie, $\nu_n \ll \infty$)\footnote{While the inferred $\nu_n = 17$ is relatively large, leading to $95\%$ and $99\%$ credibility regions that are not inordinately wider than the corresponding normal ones with the same variances, the difference will be more important at the level of posterior predictions in Sec.~\ref{sec:results_full_analysis}.} and that $(n_0, E_0)$ are only weakly correlated.%
\footnote{%
\rev{Hence, no significant correlation between $(n_0, E_0)$ is present at both the posterior and prior predictive level. Note that no prior knowledge of the correlation was assumed.} Removing \rev{the potential outlier} SQMC700 from the data, the inferred marginal distributions for $(n_0, E_0)$ are still statistically consistent with those obtained for prior Set~A and Set~B, but $(n_0,E_0)$ is weakly anti-correlated, as can be seen in Appendix~\ref{app:suppl_results}.}
We quantify here the linear correlation between two random variables as follows using the Pearson correlation coefficient, 
\begin{equation} \label{eq:pearson}
\rho = \frac{\vb{C}_{12}}{\sqrt{\vb{C}_{11}\vb{C}_{22}}} \,;
\end{equation}
no ($0 \leqslant |\rho| < 0.25$), 
weak ($0.25 \leqslant |\rho| < 0.5$), 
intermediate ($0.5 \leqslant |\rho| < 0.75$), and 
strong ($0.75 \leqslant |\rho| \leqslant 1$) correlation, if $\rho \geqslant 0$, 
or anti-correlation, if $\rho < 0$.
Note that, for the considered bivariate $t$-distribution, the covariance matrix $\vb{C}$ in Eq.~\eqref{eq:pearson} is proportional to the associated scale matrix $\vPsi$, as discussed in Appendix~\ref{app:recap_distributions}.
These findings are consistent with our prior knowledge and a direct consequence of the limited number of data points.
Although our two posterior predictive distributions tend to favor somewhat lower $(n_0, E_0)$ than the original saturation box~\eqref{eq:satbox}, as supported by the data, Fig.~\ref{fig:satbox_analysis_set_A} shows that the two constraints are overall consistent with one another at the $\gtrsim 80\%$ credibility level (see also Fig.~\ref{fig:satbox_analysis_set_B} in Appendix~\ref{app:suppl_results} for Set~B).
However, only our inferred posterior predictive distribution~\eqref{eq:posterior_pred}\footnote{That is, $t_\nu(\vmu, \vPsi)$ with the hyperparameters given in Eq.~\eqref{eq:ppd_setA} for Set~A and Eq.~\eqref{eq:ppd_setB} for Set~B.}
can, and should, be evaluated at different confidence levels to indicate uncertainties in the empirical saturation point and facilitate statistically robust comparisons with, \eg, constraints from chiral EFT based on transparent model and prior assumptions.

\subsection{DFT constraints with uncertainties}
\label{sec:results_full_analysis}

We now apply the conjugate prior approach to DFT constraints with quantified uncertainties using mixture modeling, as described in Sec.~\ref{sec:stat_framework_w_uncert}. 
All constraints summarized in Table~\ref{tab:constraints} will be used, except for FSUGarnet, FSUGold2, and RMF0XY as they have been replaced by Giuliani~\etal~\cite{Giuliani:2022yna} using improved UQ for the calibration of the covariant energy-density functionals.
We also treat SV--min and TOV--min as one model class, where we uniformly decide from which one to sample in each Monte Carlo iteration.
Hence, each Monte Carlo iteration considers $n=8$ collections of DFT data points drawn from their empirical or normal distributions.
To probe the prior sensitivity of our results, we perform again the analysis for the two prior sets specified in Sec.~\ref{sec:results_satbox_analysis}.

\begin{figure}[tb]
    \centering
    \includegraphics[width=\linewidth, page=1]{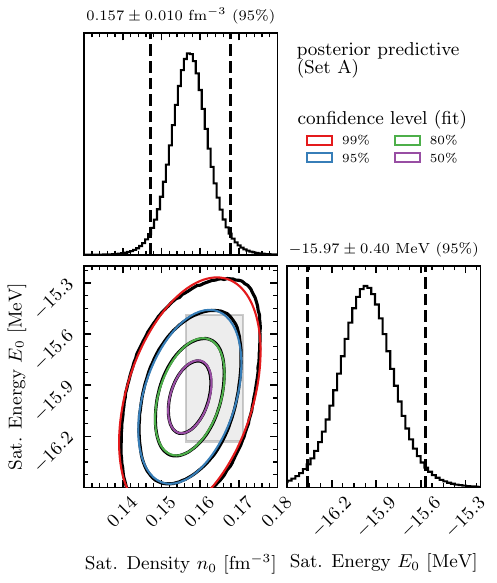}
    \caption{%
        Posterior predictive and its marginal distributions based on mixture modeling with the prior parameters in Set~A and data from Skyrme and RMF models considered.
    The colored ellipses and dashed vertical lines encompass confidence regions centered at the distributions' mean values at different credibility levels (see the legend). 
    The titles of the panels along the diagonal quote the confidence regions of the marginal distributions at the 95\% credibility level (centered at the median), corresponding to the dashed vertical lines.
    The left panel in Fig.~\ref{fig:satbox_analysis_set_A} shows the associated (data-agnostic) prior predictive.
    See the main text for details.
    } \label{fig:full_analysis_setA}
\end{figure}

For Set~A, Fig.~\ref{fig:full_analysis_setA} shows the correlation plot~\cite{corner} of the posterior predictive~\eqref{eq:full_posterior_pred}, conceptionally similarly to the right panel in Fig.~\ref{fig:satbox_analysis_set_A}.
The corresponding (data-agnostic) prior predictive is depicted in the left panel in Fig.~\ref{fig:satbox_analysis_set_A}.
We obtain the 95\% credibility regions for the two marginal distributions,
\begin{subequations} \label{eq:marginal_full_setA}
\begin{align}
    n_0^{\text{(A)}} &\approx 0.157 \pm 0.010 \fmiq \,,\\ 
    E_0^{\text{(A)}} &\approx -15.97 \pm 0.40 \MeV \,,
\end{align}
\end{subequations}
which are statistically consistent with those in Eq.~\eqref{eq:marginal_cred_range_setA} but tend to allow for somewhat lower $(n_0,E_0)$ due to the inclusion of RMF models, while their uncertainties are only slightly larger.
Random samples from the posterior predictive depicted in Fig.~\ref{fig:full_analysis_setA}, whose closed form is unknown, if it exists,\footnote{%
One might be able to derive an analytic form for the joint posterior~\eqref{eq:full_posterior_pred} after model mixing following Appendix~\ref{app:MC_for_BMM}.%
} can be straightforwardly obtained using our publicly available Jupyter notebooks~\cite{saturationGitHub}. 
However, because an analytic approximation might also be helpful, we fit a bivariate $t$-distribution to the samples from the joint posterior predictive (see Appendix~\ref{app:conf_region_bivar_t} for the technical details).
The fitted posterior predictive $t_\nu(\vmu, \vPsi)$ has $\nu_n = 9 \ll \infty$ and\footnote{One notes that the fitted degree of freedom is $\nu_n = n + 1 = 9$, with $n=8$ data points per Monte Carlo iteration.
Hence, the mixed posterior predictive~\eqref{eq:full_posterior_pred} has heavier tails than the to-be-mixed posterior predictives obtained via Monte Carlo iteration.}
\begin{equation} \label{eq:fit_bivar_t_setA}
    \vmu_n^{\text{(A)}} \approx \begin{bmatrix}
        0.157 \\ -15.97
    \end{bmatrix} \,, \quad
    \vPsi_n^{\text{(A)}} \approx \mqty[0.005^2 & 0.017^2 \\ 0.017^2 & 0.17^2] \,,
\end{equation}
indicating a weak correlation between $(n_0,E_0)$.
Colored ellipses in Fig.~\ref{fig:full_analysis_setA} encompass the confidence regions of the fitted bivariate $t$-distribution at the common confidence levels (see the legend).
They approximate well the corresponding contours (black lines) obtained from a kernel density estimation of the underlying data (\ie, without fitting to the data), demonstrating that the posterior predictive can be well represented by the fitted bivariate $t$-distribution with parameters given by Eq.~\eqref{eq:fit_bivar_t_setA}.

Appendix~\ref{app:suppl_results} has our corresponding results for Set~B. 
At the 95\% credibility level, we obtain
\begin{subequations} \label{eq:marginal_full_setB}
\begin{align}
    n_0^{\text{(B)}} &\approx 0.158 \pm 0.011 \fmiq \,,\\ 
    E_0^{\text{(B)}} &\approx -15.96 \pm 0.43 \MeV \,,
\end{align}
\end{subequations}
which are slightly more uncertain than those in Eq.~\eqref{eq:marginal_full_setA} for Set~A, as expected, yet statistically consistent, indicating only a relatively mild prior sensitivity of our results despite the data-limited case.
The fitted posterior predictive for Set~B also describes the posterior predictive reasonably well, having $\nu_n = 9$ (as for Set~A) and very similar
\begin{equation} \label{eq:fit_bivar_t_setB}
    \vmu_n^{\text{(B)}} \approx \begin{bmatrix}
        0.158 \\ -15.96
    \end{bmatrix} \,, \quad
    \vPsi_n^{\text{(B)}} \approx \mqty[0.005^2 & 0.019^2 \\ 0.019^2 & 0.19^2] \,.
\end{equation}
%


Our constraints on the saturation density of SNM are systematically larger than, although at the 95\% confidence level consistent with, the interior baryon density of \isotope[208]{Pb},  
$n_b = 0.148 \pm 0.004 \fmiq$, determined from the improved \isotope[208]{Pb} Radius EXperiment (PREX--II) at Jefferson Laboratory~\cite{Adhikari:2021phr}. 
Such a trend may be anticipated given that \isotope[208]{Pb}, a neutron-rich nucleus with an isospin asymmetry of $\delta \approx 0.21$ (as opposed to $\delta = 0$ in SNM), probes the saturation density of neutron-rich matter that is known to decrease with increasing $\delta$ (\eg, see Ref.~\cite{Piekarewicz:2008nh} and Figure~2b in Ref.~\cite{Drischler:2021kxf}).\footnote{
Note that such a trend is also reflected in most RMF models. 
Indeed, the 95\% credibility results for the RMF model obtained by Giuliani~\etal~\cite{Giuliani:2022yna} (see Fig.~\ref{fig:satpoint_constraints}) yields: $n_0 = 0.149 \ldots 0.155 \fmiq$ and $E_0 = -(16.17 \ldots 16.40) \MeV$.%
}

\subsection{Comparison with chiral EFT predictions}

\begin{figure*}[tb]
    \centering
    \includegraphics[width=0.49\linewidth,page=1]{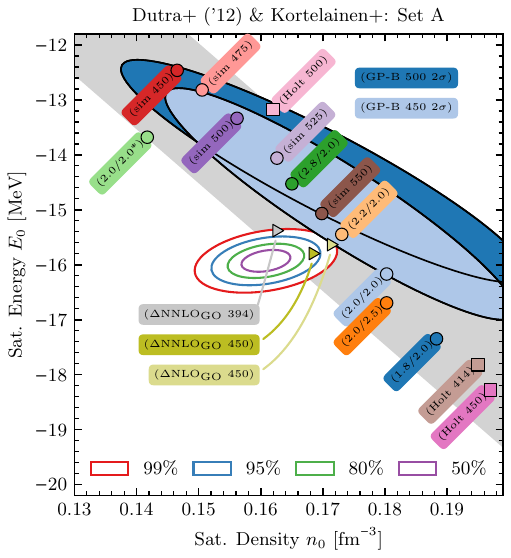}
    \includegraphics[width=0.49\linewidth,page=3]{figures/coester.pdf}
    \caption{%
    Microscopic and empirical constraints on the nuclear saturation point in comparison.
    Predictions by many-body perturbation theory (MBPT; circles, squares) and coupled cluster theory (CC; triangles) with chiral NN and 3N interactions~\cite{Hebeler:2010xb,Sammarruca:2014zia,Holt:2016pjb,Drischler:2017wtt,Jiang:2020the,Drischler:2020yad} are shown;
    the annotations specify the nuclear interactions: ``$\lambda/\Lambda_{\rm 3N} \, [\fmi]$'' for Hebeler~\etal~\cite{Hebeler:2010xb}, 
    ``sim~$\Lambda \, [\MeV]$'' for Carlsson~\etal~\cite{Carlsson:2015vda}, ``Holt~$\Lambda \, [\MeV]$'' for Holt~\etal~\cite{Holt:2016pjb,Sammarruca:2014zia}, and ``$\Delta\text{NLO}_\text{GO}~\Lambda$~[MeV]'' and ``$\Delta\text{NNLO}_\text{GO}~\Lambda$~[MeV]''~\cite{Jiang:2020the} (with explicit delta isobars), where $(\Lambda_{\rm 3N},\Lambda)$ are momentum cutoffs and $\lambda$ is the resolution scale associated with the Similarity Renormalization Group (SRG) evolution. 
    The blue ellipses (``GP--B~$\Lambda$ [MeV]'') show the $2\sigma$ (\ie, $86\%$) confidence regions of the MBPT calculations up to N$^3$LO in Ref.~\cite{Drischler:2017wtt} with to-all-orders EFT truncation errors quantified~\cite{Drischler:2020yad}. 
    The Coester-like anti-correlation band~\cite{Drischler:2015eba}, obtained from a simple linear fit to the chiral EFT data, is illustrated in gray for guidance.
    The colored ellipses with white fillings (for easier readability) encompass our inferred regions at several confidence levels (see the captions) respectively obtained in Sec.~\ref{sec:results_satbox_analysis} (left panel) and in Sec.~\ref{sec:results_full_analysis} (right panel).
    Note that the credibility regions associated with 95\% (blue ellipse) and 99\% confidence (red ellipse) do \emph{not} correspond to the traditional $2\sigma$ and $3\sigma$ regions for these heavy-tailed (bivariate) $t$-distributions.
    Modified from Figure~2b in Ref.~\cite{Drischler:2021kxf}.
    } \label{fig:eft_satpoints}
\end{figure*}

Figure~\ref{fig:eft_satpoints} compares our empirical constraints with the predicted saturation points of a wide range of chiral NN and 3N interactions (see the annotations and caption for details).
The empirical constraints inferred in Sec.~\ref{sec:results_satbox_analysis} (left panel) and Sec.~\ref{sec:results_full_analysis} (right panel), respectively, for prior Set~A are depicted by the colored ellipses with white fillings (for easier readability), which encompass the credibility levels at the common percentiles (see the legends).
Recall that these constraints are heavy-tailed bivariate $t$-distributions.
The results for prior Set~B are statistically consistent and can be found in Appendix~\ref{app:suppl_results}.
Circles and squares depict results obtained in many-body perturbation theory (MBPT) calculations, whereas the triangles were obtained in coupled cluster (CC) theory calculations.
Depicting normal distribution, the two blue ellipses correspond to the $2\sigma$ (\ie, $86\%$)\footnote{Note that confidence levels of bivariate normal distributions do not follow the well-known 68--95--99.7 rule for univariate normal distributions. See the discussion of Eq.~\eqref{eq:normal_regions} in Appendix~\ref{app:conf_region_bivar_t}.} confidence regions of the N$^3$LO calculations in Ref.~\cite{Drischler:2017wtt} with to-all-orders EFT truncation errors quantified~\cite{Drischler:2020yad,Drischler:2020hwi} (\ie, ``GP--B~500'' and ``GP--B~450''). 
For a detailed discussion of recent EFT predictions of the nuclear saturation point, we refer the reader to, \eg, Refs.~\cite{Drischler:2021kxf,Machleidt:2024bwl}.

As shown in Fig.~\ref{fig:eft_satpoints}, only one chiral Hamiltonian (with explicit delta isobars) saturates within the 95\% confidence region of the inferred saturation points.
However, rigorously ascertaining the agreement (or disagreement) of these point estimates would require quantified theoretical uncertainties in the microscopic predictions, including estimates of the EFT truncation error.
The ``GP--B~450'' confidence ellipses, which quantify these EFT truncation errors, barely overlap with our 99\% confidence regions despite its sizable uncertainties (at the $2\sigma$ confidence level), while most of the other Hamiltonians saturate significantly outside this confidence region.
However, the chiral interactions labeled ``Holt~500,'' ``sim~500'' and ``sim~520,'' ``$\Delta\text{NNLO}_\text{GO}~394$,'' and ``2.8/2.0'' (see the caption of Fig.~\ref{fig:eft_satpoints} for details) saturate at densities $n_0$ consistent with the inferred empirical ranges for the saturation density.
On the other hand, the chiral interactions labeled ``2.0/2.0,'' ``2.2/2.0,'' and the three deltaful ``GO'' potentials saturate consistent with the empirical ranges for the saturation energy.
In conclusion, we find (consistent with the literature) that the various microscopic SNM calculations depicted in Fig.~\ref{fig:eft_satpoints} are challenged by simultaneously predicting $(n_0,E_0)$ both accurately and, within the uncertainties, in agreement with our inferred empirical constraints (see also the similar discussions in Refs.~\cite{Drischler:2020yad,Drischler:2020hwi,Drischler:2021kxf}). 

\subsection{Nuclear symmetry energy constraints}

Next, we study the implications of the inferred empirical saturation point on the nuclear symmetry energy.
In the standard quadratic approximation of the EOS's isospin asymmetry ($\delta$) dependence, the symmetry energy at a given baryon number (density) $\nb$ is determined by the difference between the energies per particle in PNM ($E/N$, $\delta=1$) and SNM ($E/A$, $\delta=0$) (see also, \eg, Refs.~\cite{Lattimer:2023rpe,Drischler:2021kxf}),
\begin{align}
	S(\nb) &\approx \frac{E}{N}(\nb) - \frac{E}{A}(\nb) \,, \label{eq:S(\nb)} \\ 
	&\equiv S_v + L \left( \frac{\nb-n_0}{3n_0} \right) + \ldots \, . \label{eq:S(\nb)_taylored}
\end{align}
Here, we focus on the symmetry energy parameters $(S_v,L)$ in the series expansion~\eqref{eq:S(\nb)_taylored} of $S(\nb)$ about $n_0$:
\begin{equation} \label{eq:Sv_L_def}
    S_v = S(n_0) \quad \text{and} \quad 
    L = 3 n_0 \, \frac{\dd}{\dd{n}} S(\nb)\bigg|_{\nb=n_0} \,.
\end{equation}
These correspond to the symmetry energy evaluated at $n_0$ and its so-called slope parameter, respectively.
Because $E(\nb)/A$ in SNM exhibits a minimum at $n_0$ and we defined $E_0\equiv E(n_0)/A$, the symmetry energy parameters~\eqref{eq:Sv_L_def} can be expressed as
\begin{align} \label{eq:Sv_L_E/N}
	S_v = \frac{E}{N}(n_0) - E_0 \quad \text{and} \quad 
	 L = \frac{3}{n_0} p(n_0)\,,
\end{align}
with the PNM energy per particle $E(n_0)/N$ and pressure 
\begin{equation} \label{eq:pressure}
    p(n_0) = n_0^2 \, \dv{\nb} \frac{E}{N}(\nb) \bigg|_{\nb=n_0} \,,
\end{equation}
evaluated at the saturation density~$n_0$. 
Note that the definition of the  pressure~\eqref{eq:pressure} is not to be confused with the probability distribution $P(n_0)$.

To constrain the symmetry energy parameters~\eqref{eq:Sv_L_E/N}, we choose here the PNM calculations at N$^3$LO conducted in Ref.~\cite{Drischler:2017wtt} for two momentum cutoffs ($\Lambda = 500$ and $450 \MeV$) and with (correlated) to-all-orders EFT truncation errors quantified~\cite{Drischler:2020yad,Drischler:2020hwi}. 
We combine these with the (tightly constrained) empirical constraints for $(n_0,E_0)$ in SNM instead of the corresponding microscopic calculations.
Theoretical uncertainties in microscopic PNM calculations (at that density) are generally more controlled than they are in SNM (\eg, those depicted in Fig.~\ref{fig:eft_satpoints}) since the three-neutron forces (in PNM) are overall weaker and, through N$^3$LO in Weinberg power counting, entirely determined by the NN potential.
Hence, microscopic predictions for $E(\nb)/N$ and $p(\nb)$ at $\nb \lesssim n_0$ obtained from different many-body frameworks and with different chiral interactions typically agree within their uncertainties, \eg, as Figure~1 in Ref.~\cite{Huth:2020ozf} shows.
Other recent nuclear matter calculations based on microscopic NN and 3N interactions, including constraints on the symmetry energy, can be found in, \eg, Refs.~\cite{Lonardoni:2019ypg,Wen:2020nqs,Hu:2021trw,Jiang:2022oba,Jiang:2022tzf}.

Specifically, we use the PNM EOSs labeled ``GP--B~500~[\text{MeV}]'' and ``GP--B~450~[\text{MeV}],'' which are represented by GPs and provided by the BUQEYE collaboration~\cite{BUQEYEweb} through publicly available Jupyter notebooks~\cite{BUQEYEsoftware}.
Gaussian Processes are used for interpolating and quantifying the EFT truncation errors in $E(n)/N$ and $p(n)$ in PNM, as discussed in Refs.~\cite{Drischler:2020yad,Drischler:2020hwi}.
We propagate the uncertainties in the inferred empirical $(n_0,E_0)$ and provided microscopic $(E(n_0)/N,p(n_0))$ to $(S_v,L)$ via marginalization, 
\begin{equation} \label{eq:joint_sv_l}
    P(S_v,L) = \int \dd{n_0} \int \dd{E_0} P(S_v,L \mid n_0, E_0) \, P(n_0, E_0) \,,
\end{equation}
where $P(n_0, E_0)$ denotes the inferred posterior predictive~\eqref{eq:full_posterior_pred} [\ie, a bivariate $t$-distribution] and $P(S_v,L \mid n_0, E_0)$ the joint distribution of $(S_v,L)$ as obtained from Eqs.~\eqref{eq:Sv_L_E/N} and~\eqref{eq:pressure} for a given $(n_0, E_0)$.
We account for correlations between $E(n_0)/N$ and $p(n_0)$ but assume that both are uncorrelated with the empirical $(n_0,E_0)$, as they were derived from distinct frameworks (\ie, microscopic EFT vs. phenomenological DFT). 
If $(n_0,E_0)$ were instead also predictions from the same chiral Hamiltonian, $E(n_0)/N$, $p(n_0)$, and $(n_0,E_0)$ would be correlated (see Refs.~\cite{Drischler:2020yad,Drischler:2020hwi} for details).
For similar analyses of $(S_v, L)$ constrained by microscopic PNM calculations and the empirical saturation point, we refer the reader to, \eg, Refs.~\cite{Hebeler:2010jx,Gandolfi:2011xu,Tews:2012fj,Holt:2016pjb}.

With BUQEYE's Jupyter notebooks~\cite{BUQEYEsoftware}, sampling the joint (bivariate normal) distribution $P(E(n_0)/N, p(n_0))$, and thus via Eq.~\eqref{eq:Sv_L_E/N} also $P(S_v,L \mid n_0, E_0)$, is straightforward.
Following a similar two-step process as described in Sec.~\ref{sec:results_full_analysis}, we first draw $M=10^4$ random samples from $P(n_0, E_0)$ and then, for each of those points, we sample $N=800$ points from $P(S_v,L \mid n_0, E_0)$, resulting in a total of $Q = M\times N = 8\times10^6$ sampling points of $P(S_v,L)$ given by Eq.~\eqref{eq:joint_sv_l}.
For simplicity, we approximate $P(S_v,L) \approx \mathcal{N}_2(\vmu, \vSigma)$ using the sample mean $\vmu$ and sample covariance $\vSigma$ to obtain a closed form of the distribution.\footnote{Fitting bivariate $t_\nu(\vmu, \vPsi)$ to the samples resulted in approximately normal distributions due to large degrees of freedoms $\nu$.}
For the results in Sec.~\ref{sec:results_satbox_analysis} with Set~A and ``GP--B 500,'' we find
\begin{equation} \label{eq:satbox_sv_l_set_a_500}
    \vmu^{\text{(A)}} \equiv \begin{bmatrix}
    S_v\\
    L
    \end{bmatrix}
    \approx \begin{bmatrix}
    32.4 \\ 54.9
    \end{bmatrix} \,,  \quad 
    \vSigma^{\text{(A)}} \approx \mqty[1.1^2 & 2.9^2 \\ 2.9^2 & 8.1^2] \,,
\end{equation}
and for ``GP--B 450''
\begin{equation} \label{eq:satbox_sv_l_set_a_450}
    \vmu^{\text{(A)}} \approx \begin{bmatrix}
    33.5 \\ 61.0
    \end{bmatrix} \,, \quad 
    \vSigma^{\text{(A)}} \approx \mqty[1.0^2 & 2.6^2 \\ 2.6^2 & 7.2^2] \,,
\end{equation}
where we have omitted the unit~MeV for both, $S_v$ and $L$.
Likewise, for the results in Sec.~\ref{sec:results_full_analysis} with Set~A and ``GP--B 500,'' we find
\begin{equation} \label{eq:full_sv_l_set_a_500}
    \vmu^{\text{(A)}} 
    \approx \begin{bmatrix}
    32.0 \\ 52.6
    \end{bmatrix} \,,  \quad 
    \vSigma^{\text{(A)}} \approx \mqty[1.1^2 & 2.9^2 \\ 2.9^2 & 8.1^2] \,,
\end{equation}
and for ``GP--B 450''
\begin{equation} \label{eq:full_sv_l_set_a_450}
    \vmu^{\text{(A)}} \approx \begin{bmatrix}
    33.1 \\ 58.6
    \end{bmatrix} \,, \quad 
    \vSigma^{\text{(A)}} \approx \mqty[1.0^2 & 2.7^2 \\ 2.7^2 & 7.4^2] \,.
\end{equation}
For Set~B, we find that the mean vectors and covariance matrices are consistent with Eqs.~\eqref{eq:satbox_sv_l_set_a_500}--\eqref{eq:full_sv_l_set_a_450} for ``GP--B 500'' and ``GP--B 450,'' respectively, and so we focus here on the results for prior Set~A only.

\begin{figure*}[tb]
    \centering
    \includegraphics[width=0.49\linewidth, page=1]{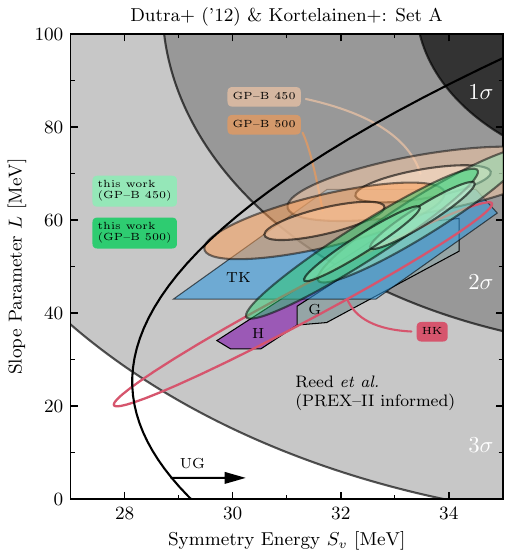}
    \includegraphics[width=0.49\linewidth, page=3]{figures/esym.pdf}
    \caption{%
    Constraints on the symmetry energy parameters $S_v$ and its slope parameter $L$ evaluated at the nuclear saturation density, and the conjectured constraint from the unitary gas (UG; see the annotations). 
    The left panel (right panel) shows the results for the saturation point inferred in Sec.~\ref{sec:results_satbox_analysis} (in Sec.~\ref{sec:results_full_analysis}) based on prior Set~A. 
    See Appendix~\ref{app:suppl_results} for the corresponding results for Set~B.
    Microscopic PNM-based constraints are given by Hebeler~\etal (H)~\cite{Hebeler:2010jx}, Holt \& Kaiser (HK; $2\sigma$ confidence level)~\cite{Holt:2016pjb},
    Gandolfi~\etal (G)~\cite{Gandolfi:2011xu}, and Tews \& Kr{\"u}ger~\etal (TK)~\cite{Tews:2012fj}.
    The orange ellipses~\cite{Drischler:2020hwi,Drischler:2020yad} depict the confidence regions from combined PNM and SNM calculations at N$^3$LO~\cite{Drischler:2017wtt} with EFT truncation errors quantified and two different sets of chiral NN and 3N interactions (light: $1\sigma$ (39\%), dark: $2\sigma$ (86\%)). 
    The three ellipses (black: $1\sigma$ (39\%), dark gray: $2\sigma$ (86\%), light gray: $3\sigma$ (99\%)) in the background correspond to the PREX--II informed constraint obtained in Ref.~\cite{Reed:2021nqk} based on covariant energy functionals.
    Modified and extended from Figure~4b in Ref.~\cite{Drischler:2021kxf}.
    } \label{fig:esym}
\end{figure*}

Figure~\ref{fig:esym} depicts our constraints on $(S_v,L)$ by the green confidence ellipses (light: $1\sigma$ (39\%), dark: $2\sigma$ (86\%)).
The left panel (right panel) shows the results for the saturation point inferred in Sec.~\ref{sec:results_satbox_analysis} (in Sec.~\ref{sec:results_full_analysis}) for prior Set~A. 
See Appendix~\ref{app:suppl_results} for the corresponding results for Set~B.
The two orange confidence ellipses (light: $1\sigma$ (39\%), dark: $2\sigma$ (86\%)) were derived in Refs.~\cite{Drischler:2020hwi,Drischler:2020yad} based on the same quadratic approximation~\eqref{eq:Sv_L_E/N} and microscopic PNM calculations we study in this work. 
But instead of the empirical $(n_0,E_0)$, Refs.~\cite{Drischler:2020hwi,Drischler:2020yad} considered the saturation point predicted by the underlying chiral Hamiltonians and computed the slope parameter via the density derivative in Eq.~\eqref{eq:Sv_L_def} while accounting for correlations in the predicted $(n_0,E_0)$ in SNM and $(E(n_0)/N,p(n_0))$.
As one can see in Fig.~\ref{fig:eft_satpoints}, these predicted saturation points (blue ellipses) are systematically shifted towards higher $(n_0,E_0)$ relative to the inferred empirical saturation points, which propagates to the symmetry energy predictions (\eg, larger $L$ values).
However, unlike our semi-microscopic approach, the fully microscopic predictions in Refs.~\cite{Drischler:2020hwi,Drischler:2020yad} can exploit correlations between the microscopic predictions for $E(n_0)/N$ and $E_0$ to constrain the symmetry energy parameters~\eqref{eq:Sv_L_E/N} more tightly than the in-quadrature sum of the individual uncertainties of $E(n_0)/N$ and $E_0$.

Figure~\ref{fig:esym} also compares our results with similar PNM-based constraints by Hebeler~\etal (H)~\cite{Hebeler:2010jx}, Holt \& Kaiser (HK)~\cite{Holt:2016pjb}, Gandolfi~\etal (G)~\cite{Gandolfi:2011xu}, and Tews \& Kr{\"u}ger~\etal (TK)~\cite{Tews:2012fj}.
The allowed region associated with the conjectured unitary gas (UG) constraint is depicted by the solid black line (in the direction of the black arrow).
In addition, Fig.~\ref{fig:esym} shows the PREX--II-informed constraint at the $1\sigma$ (black; 39\%), $2\sigma$ (dark gray; 86\%), and $3\sigma$ (light gray; 99\%) confidence level recently derived in Ref.~\cite{Reed:2021nqk} from 16 covariant energy density functionals, including FSUGold2.
This empirical constraint is given by
$(S_v, L) \sim \mathcal{N}_2(\vmu, \vSigma)$ with
\begin{equation} \label{eq:Sv_L_PREX}  
    \vmu = \begin{bmatrix} 
    38.1\\
    106
    \end{bmatrix}  \quad \text{and} \quad
    \vSigma = \mqty[\dmat[0]{4.7^2, 37^2}] \,,
\end{equation}
where we constructed the covariance matrix $\vSigma$ from the two marginal distributions given in Ref.~\cite{Reed:2021nqk}, assuming that the PREX--II-informed $(S_v,L)$ are uncorrelated to be conservative with UQ.
In fact, Ref.~\cite{Reed:2021nqk} predicts $(S_v,L)$ to be strongly correlated ($\rho \approx 1$), which is not manifested in the marginal distributions and
likely underestimates the uncertainties in the $(S_v,L)$ plane.
Note that the mean vector in Eq.~\eqref{eq:Sv_L_PREX} (\ie, the mode of the distribution) is outside the $(S_v,L)$ range depicted in Fig.~\ref{fig:esym} and inconsistent with the conjectured UG constraint.
Also, note that the $1\sigma$ region of bivariate normal distributions corresponds to the relatively low confidence level of 39.4\% (see Appendix~\ref{app:conf_region_bivar_t}). 
More detailed discussions of the implications of PREX--II at Jefferson Laboratory~\cite{Adhikari:2021phr} on the EOS can be found, \eg, in Refs.~\cite{Xu:2020fdc,Piekarewicz:2021jte,Essick:2021ezp,Essick:2021kjb,Reinhard:2021utv,Reinhard:2022inh,Arthuis:2024mnl}.

As shown in Fig.~\ref{fig:esym}, our marginalized constraints on $S_v$ are consistent with the previous PNM-based calculations.
The strong correlation between $(S_v,L)$ we obtain agrees well with those of the other PNM-based calculations.
However, except for the other N$^3$LO calculation labeled ``TK,'' our inferred $L$ values are systematically larger than the constraints ``H,'' ``G,'' and ``HK,'' consistent with the finding that the PNM EOSs used in this work are relatively stiff at densities $\nb \approx n_0$~\cite{Drischler:2021bup}.
On the other hand, the microscopic (SNM-informed) constraints obtained in Ref.~\cite{Drischler:2020hwi,Drischler:2020yad} (orange ellipses), which account for correlations between $E(n_0)/N$ and $E_0$ when computing the symmetry energy~\eqref{eq:Sv_L_E/N}, predict a weaker correlation between $(S_v, L)$ and slightly larger $L$ values.
Our constraints for the two momentum cutoffs are statistically consistent with one another (at the $1\sigma$ level), as expected from the relatively mild cutoff dependence observed in Figure~3 in Ref.~\cite{Drischler:2017wtt}.

All microscopic calculations (at the confidence level shown) are statistically consistent with the PREX--II-informed constraint~\eqref{eq:Sv_L_PREX} at the $2\sigma$ (\ie, 86\%) confidence level or better. 
They fall completely within its $3\sigma$ (\ie, 99\%)  region except for the lower tail of the microscopic constraint ``HK.''
However, the PREX--II-informed constraint~\eqref{eq:Sv_L_PREX} favors larger $L$ values with significant uncertainties.
The anticipated $\pm 0.03$~fm precision on the neutron skin thickness of \isotope[208]{Pb} by the Mainz Radius Experiment (MREX) at the Mainz Energy-Recovering Superconducting Accelerator (MESA) will be critical in narrowing the experimentally informed constraints on $(S_v,L)$~\cite{Mammei:2023kdf}.
\section{Summary and Outlook} 
\label{sec:summary_outlook}

Density functional theory provides important empirical constraints for benchmarking in-medium nuclear interactions and modeling the nuclear EOS. One fundamental constraint is the equilibrium density of SNM ($n_{0}$) which represents the density at which the pressure vanishes and the energy per particle $E_0$ exhibits its minimum value. Collectively, this is referred to as the nuclear saturation point $(n_0, E_0)$.
As shown in Fig.~\ref{fig:satpoint_constraints}, the various DFT predictions \rev{from Skyrme and RMF models} considered here report $(n_0, E_0)$ with high precision 
(or unspecified precision in cases without UQ) leading to inconsistencies across multiple model predictions at high confidence levels.\footnote{This inconsistency between to-be-mixed model predictions distinguishes our work from, \eg, the Bayesian model mixing approach to the SNM EOS presented in Ref.~\cite{Semposki:2024vnp}.}
In essence, the DFT constraints incorporated in this work---all of them informed by nuclear observables---cannot all be simultaneously precise \emph{and} accurate in the determination of the saturation point. 
The inconsistency is also reflected by the well-known trend that Skyrme models systematically predict a higher saturation density than RMF models. 
Hence some, if not all, models, necessarily feature UQ inaccuracies due to unknown modeling uncertainties that are generally difficult to quantify. One hard to quantify model uncertainty is the surface energy, a critical component to the energy budget of finite nuclei but irrelevant to the study of infinite nuclear matter. In particular, a softer surface energy can compensate for a higher saturation density, ultimately leading to the same charge radii in both Skyrme and RMF models.  These issues motivate the question of how these empirical constraints from DFT can be leveraged to benchmark saturation properties of chiral interactions.

In this work, we proposed a Bayesian hierarchical model that estimates the true
empirical saturation point by mixing multiple DFT constraints.
The model assumption is that each DFT constraint represents a random sample from a universal (bivariate) normal distribution modeling the unobserved true saturation, whose mean vector and covariance matrix are to be inferred from the data. 
As detailed in Sec.~\ref{sec:stat_framework}, our Bayesian framework is computationally efficient due to distributional conjugacy and, if uncertainties in the DFT constraints were estimated, ordinary Monte Carlo sampling with variance reduction is sufficient to sample from the posterior distributions. 
The resulting posterior predictive distributions for $(n_0, E_0)$ \rev{are} in the form of mixtures of correlated, bivariate student $t$-distributions, where the mixing distributions are given by the DFT developers' stated uncertainties. 
The model parameters (\ie, the bivariate mean vector and covariance matrix of the
true
saturation point) are also treated as mixtures, with respect to these uncertainties, having the Normal-inverse-Wishart (NIW) distribution.
Their confidence regions (\ie, ellipses) can be analytically determined and conveniently plotted at given credibility levels, as discussed in Appendix~\ref{app:conf_region_bivar_t}.\footnote{Specifically, our efficient posterior computation and mathematically described confidence ellipses are due to our normal likelihood model with NIW prior on its parameters, which results in NIW posteriors on these parameters and multivariate posterior predictive $t$-distributions. 
Using mixtures to incorporate DFT models' uncertainties preserves this convenience, requiring only the further step of ordinary Monte Carlo sampling.}
The associated marginal distributions are univariate $t$-distributions with the same updated degree of freedom.
Our Bayesian framework is publicly available~\cite{saturationGitHub} so that practitioners can readily use and extend our results in their work. 
It generally applies to mixing multivariate distributions, not just bivariate DFT constraints for the nuclear saturation point.


\begin{table}[tbp]
    \centering
    \caption{Summary of the central results: marginalized $95\%$-level constraints on the empirical saturation point $(n_0, E_0)$, nuclear symmetry energy $S_v$, and slope parameter $L$ evaluated at $n_0$ for the two scenarios discussed in Secs.~\ref{sec:results_satbox_analysis} (``Skyrme only'') and~\ref{sec:results_full_analysis} (``Skyrme + RMF''), respectively. 
    Only the results for prior Set~A [Eq.~\eqref{eq:prior_setA}] and the PNM EOS ``GP--B 500'' are considered.
    The units are $\MeV$ except for $n_0$, which is given in $\fmiq$.
    We stress that the joint distributions for $(n_0, E_0)$ and $(S_v,L)$ are (correlated) bivariate $t$- and normal distributions, respectively, whose parameters are referenced in the footnotes.}
    \label{tab:summary_table}
        \begin{ruledtabular}
    \begin{tabular}{ScScd{2.4}d{2.4}}
         \multicolumn{1}{Sc}{} & \multicolumn{1}{Sc}{Distribution} & \multicolumn{1}{Sc}{Skyrme only} & \multicolumn{1}{Sc}{Skyrme + RMF}\\
        \hline
        $n_0$ & $t_\nu(\mu,\Psi)$ & 0.161(7)\footnotemark[1] & 0.157(10)\footnotemark[2] \\
        $E_0$ & $t_\nu(\mu,\Psi)$ & -15.93(35)\footnotemark[1] & -15.97(40)\footnotemark[2] \\
        $S_v$ & $\mathcal{N}(\mu,\Sigma)$ & 32.4(1.1)\footnotemark[3] & 32.0(1.1)\footnotemark[4] \\
        $L$ & $\mathcal{N}(\mu,\Sigma)$ & 54.9(8.1)\footnotemark[3] & 52.6(8.1)\footnotemark[4]
    \end{tabular}
     \end{ruledtabular}
        \footnotetext[1]{See Eq.~\eqref{eq:ppd_setA} for the corresponding bivariate $t_{\nu}(\vmu, \vPsi)$.}
        \footnotetext[2]{See Eq.~\eqref{eq:fit_bivar_t_setA} for the corresponding (fitted) bivariate $t_{\nu}(\vmu, \vPsi)$.}
        \footnotetext[3]{See Eq.~\eqref{eq:satbox_sv_l_set_a_500} for the corresponding bivariate $\mathcal{N}_2(\vmu, \vSigma)$.}
        \footnotetext[4]{See Eq.~\eqref{eq:full_sv_l_set_a_500} for the corresponding bivariate $\mathcal{N}_2(\vmu, \vSigma)$.}
\end{table}

In Sec.~\ref{sec:results_discussion}, we applied this Bayesian framework to two collections of DFT constraints (see also Fig.~\ref{fig:satpoint_constraints}).
Table~\ref{tab:summary_table} summarizes the central results.
The first collection was used in Ref.~\cite{Drischler:2015eba} to construct the saturation box~\eqref{eq:satbox} depicted in Figs.~\ref{fig:satpoint_constraints} and~\ref{fig:satbox} based on the range predicted by 14 Skyrme models without considering UQ.
It was subsequently slightly extended and applied to benchmark nuclear saturation properties of chiral interactions~\cite{Drischler:2017wtt}.
In that case, our mixture model simplifies to a conjugate distribution approach without needing Monte Carlo sampling.
The second collection consisted of various DFT constraints, including RMF and Skyrme models, with and without UQ.
The inferred posterior predictive distributions for $(n_0,E_0)$ are generally weakly correlated, bivariate $t$-distributions with heavy tails.
Despite the data-limited scenario, we found only a relatively mild prior sensitivity based on calculations with two prior sets.
Our findings are consistent with the original saturation box~\eqref{eq:satbox} at the $\gtrsim 80\%$ credibility level, although they allow for somewhat lower $(n_0, E_0)$.
However, we recommend that our posterior predictive be evaluated at multiple confidence levels (see also Appendix~\ref{app:conf_region_bivar_t}) to indicate uncertainties in the empirical saturation point and facilitate statistically robust comparisons with, \eg, predictions from chiral EFT.
None of the nuclear interactions we considered here saturated within the 80\% confidence region of the inferred empirical saturation point.

We also constrained the symmetry energy $S_v$ and its density dependence $L$ evaluated at $n_0$ in the standard quadratic approximation of the EOS's isospin dependence (see Table~\ref{tab:summary_table} for the results).
To this end, we combined the inferred empirical saturation point in SNM with recent microscopic calculations of the energy per particle in PNM, where chiral NN and 3N interactions were included at N$^3$LO and correlated, to-all-orders EFT truncation errors quantified.
Our results for $(S_v,L)$, including their correlations, are consistent with similar, previous PNM-based constraints, especially with the N$^3$LO constraints obtained in Ref.~\cite{Tews:2012fj}. 
The recent PREX--II-informed constraint~\cite{Reed:2021nqk} is reproduced at the $2\sigma$ (\ie, 86\%) confidence level. 

Our constraints on the empirical saturation point enable more rigorous benchmarks of microscopic interactions derived from deltaless and deltaful EFT.
The relative agreement between competing EFT predictions and the inferred empirical saturation point can be quantified, \eg, using the Jenssen--Shannon or Jeffreys distance.
These distances could be used to calibrate chiral NN and 3N interactions to the empirical saturation point, facilitated by fast and accurate emulators for nuclear matter~\cite{Jiang:2022tzf,Jiang:2022oba}.
Such rigorous in-medium benchmarks may lead to predictive, microscopic interactions for state-of-the-art nuclear structure calculations up to heavy nuclei and infinite nuclear matter calculations for studying neutron stars.
Conversely, our empirical constraints may guide the development of Skyrme models that incorporate microscopic physics (\ie, long-range pion interactions) via the density matrix expansion (DME)~\cite{Dyhdalo:2016tgx,Zurek:2020zys,Zurek:2023mdh}, where the model parameters are constrained by $(n_0,E_0)$ and other low-density EOS properties.

Future applications of our Bayesian framework will need to quantify and reduce multicollinearity in the to-be-mixed DFT predictions for $(n_0,E_0)$. 
That is, DFT models similar in their functional forms and/or parameter estimation protocol should not be assigned the same mixing weights as noncollinear DFT models. 
Frameworks for addressing model collinearities would also benefit, in general, other model mixing efforts such as, for example, those involving nuclear masses~\cite{Neufcourt:2019sle,Neufcourt:2020nme,Kejzlar:2023tlm}. 
However, we stress that this will likely not alleviate the systematic discrepancy between the Skyrme and RMF models depicted in Fig.~\ref{fig:satpoint_constraints}, which is mainly responsible for the uncertainties in the inferred saturation point.
Elucidating the origin of this discrepancy remains an important task for nuclear DFT.
Another research avenue will be Bayesian UQ and model mixing of different DFT models directly, involving both Skyrme and RMF models, rather than just their predictions for the nuclear saturation point. 
These research avenues are currently pursued by the Bayesian Analysis of Nuclear Dynamics (BAND) collaboration~\cite{Phillips:2020dmw,bandframework}.
Parity-violating electron scattering experiments, \eg, at the upcoming facility MESA~\cite{Mammei:2023kdf}, will provide exciting electroweak probes of the interior densities of heavy nuclei to validate and improve these next-generation theoretical models and their uncertainties.

\begin{acknowledgments}

We thank S.~K.~Bogner, \rev{B.~}A.~Brown, R.~J.~Furnstahl, K.~Godbey, M.~Grosskopf, C.~J.~Horowitz, W.~Nazarewicz, D.~R.~Phillips\rev{, and P.~G. Reinhard} for fruitful discussions and the BAND collaboration~\cite{BAND_Framework,bandframework} for encouragement.
C.D. is grateful to F.~M.~Nunes for connecting him with the Summer Research Opportunities Program (SROP)~\cite{srop} at Michigan State University (MSU), under which parts of this research were conducted.
S.B. thanks the MSU SROP faculty and FRIB for the opportunity to conduct undergraduate research under the SROP program. 
This material is based upon work supported by the U.S. Department of Energy, Office of Science, Office of Nuclear Physics, under the FRIB Theory Alliance Award No. DE-SC0013617, under the STREAMLINE Collaboration awards DE-SC0024646 (Florida State University), DE-SC0024586 (Michigan State University), and DE-SC0024233 (Ohio University), and under Award No. DE-FG02-92ER40750 (J.P.). F.V. gratefully acknowledges the support of NSF from awards \rev{OAC}-2004601 \rev{(BAND Collaboration)} and DMS-2311306 for this research.
\end{acknowledgments}

\appendix
\section{Probability distribution functions}
\label{app:recap_distributions}

We give here a brief, nonexhaustive overview of the probability distribution functions used in this work and some of their useful properties. 
For more information, we refer the reader to, \eg, Ref.~\cite{gelman2013bayesian}.

\subsection{Inverse-Wishart and Normal-inverse-Wishart distribution}
\label{app:IW-NIW}

The inverse-Wishart (IW) distribution $\mathcal{W}_d^{-1}(\nu, \vb*{\Psi})$ is a multivariate generalization of the scaled inverse-$\chi^2$ distribution. 
For fixed positive integer $d$, the distribution $\mathcal{W}_d^{-1}(\nu, \vb*{\Psi})$ is supported by the set of $d\times d$ symmetric, positive-definite matrices, and its parameters are the degree-of-freedom integer number $\nu > d-1$ and the $d\times d$ scale matrix $\vb*{\Psi}$. 
The scale matrix has to be real-valued, symmetric, and positive-definite.
Random symmetric, positive-definite matrices $\vb{X} \sim \mathcal{W}^{-1}_d(\vb{X} \mid\nu, \vb*{\Psi})$ have the probability density\footnote{%
The inverse of the random variable has the so-called Wishart distribution $\mathcal{W}_d$: $\vb{X}^{-1} \sim \mathcal{W}_d(\nu, \vb*{\Psi}^{-1})$.}
\begin{multline} \label{eq:pdf_invwish}
        \mathcal{W}^{-1}_d
        (\vb{X}\mid\nu, \vb*{\Psi}) = \frac{|\vb*{\Psi}|^\frac{\nu}{2}}{2^{ \frac{\nu d}{2} }
           |\vb{X}|^{\frac{\nu + d + 1}{2}} \Gamma_d \left(\frac{\nu}{2} \right)}\\
           \times \exp\left[ -\frac{1}{2}\tr(\vb*{\Psi} \vb{X}^{-1}) \right] \,,
\end{multline}
where $\Gamma_d(x)$ denotes the $d$-variate gamma function, \mbox{$|\bullet|$} is the usual determinant functional, and $\tr(\bullet)$ is the usual trace operator, on the space of $d\times d$ matrices.
The distribution's mean value and mode are respectively
\begin{align}
    \operatorname{E}[\vx] &= \frac{\vb*{\Psi}}{\nu - d -1} \qc \text{for} \quad \nu > d+1 \,, \label{eq:mean_val_invwish}\\
    \operatorname{Mo}[\vx] &= \frac{\vb*{\Psi}}{\nu + d +1} \,.
\end{align}
Note that, in the bivariate case (\ie, $d=2$) considered in this work, the mean~\eqref{eq:mean_val_invwish} is only well-defined if $\nu \geqslant 4$.    
The inverse-Wishart distribution is implemented in the Python package \texttt{scipy.stats} as \texttt{invwishart}~\cite{2020SciPy-NMeth}.

Now, let us suppose that a random two-dimensional vector $\vmu$ is distributed as
\begin{equation}
\vmu \mid \vmu_0,\kappa,\vSigma \sim \mathcal{N}_d\left(\vb*{\mu} 
\mid \vmu_0,\frac{1}{\kappa}\vSigma\right),
\end{equation}
that is, a multivariate normal distribution with mean $\vmu_0$ and covariance matrix $\frac{1}{\kappa}\vSigma$, where $\kappa$ is a positive integer. Next, suppose that, independently of the randomness in the vector $\vmu$, its parameter $\vSigma$ is also random, and specifically
\begin{equation}
    \vSigma \mid \boldsymbol\Psi,\nu \sim \mathcal{W}_d^{-1}(\vSigma \mid \nu, \boldsymbol\Psi)
\end{equation}
is inverse-Wishart distributed with degree of freedom $\nu$ and scale matrix $\vb*{\Psi}$. 
Then, the joint distribution of the pair of variables $(\vmu,\vSigma)$
is known as a Normal-inverse-Wishart (NIW) distribution, denoted as
$(\vmu,\vSigma) \sim \operatorname{NIW}(\vmu,\vSigma \mid \vmu_0,\kappa,\boldsymbol\Psi,\nu)$. 
This construction leads directly to the probability density function of the pair $(\vmu,\vSigma)$:
\begin{align}
    \operatorname{NIW}(\vmu,&\vSigma \mid \vmu_0,\kappa,\boldsymbol\Psi,\nu) \notag \\
    &= \mathcal{N}_d\left(\vmu \mid \vmu_0,\frac{1}{\kappa}\vSigma\right) \mathcal{W}_d^{-1}(\vSigma \mid \boldsymbol\Psi,\nu)\\
&=\frac{\kappa^{d/2}|\boldsymbol{\Psi}|^{\nu /
    2}|\boldsymbol{\Sigma}|^{-\frac{\nu + d + 2}{2}}}{(2
  \pi)^{d/2}2^{\frac{\nu d}{2}}\Gamma_d(\frac{\nu}{2})} \exp\bigg[
  -\frac{1}{2} \tr(\boldsymbol{\Psi
    \Sigma}^{-1}) \notag \\
    & \quad -\frac{\kappa}{2}(\vmu-\vmu_0)^\intercal\boldsymbol{\Sigma}^{-1}(\vmu
  - \vmu_0) \bigg] \,. \label{eq:NIW_pdf}
\end{align}
The integer scalars $\kappa > 0$ and $\nu > d-1$, the vector $\vmu_0$, and the scale matrix $\vb*{\Psi}$ are the parameters of the NIW distribution.
Samples from the NIW distribution can be obtained in a two-step process:
First, one samples $\vSigma$ from the inverse-Wishart distribution with scale matrix $\vb*{\Psi}$ and $\nu$ degree of freedom.
Then, one samples $\vb*{\mu}$ from a multivariate normal distribution with mean $ {\vmu}_{0}$ and covariance $\frac {1}{\kappa } \vSigma$.
One repeats the process until the desired number of sampling points, \ie, $\{(\vmu_i,\vSigma_i)\}_i$, is obtained.

\subsection{Multivariate student $t$-distribution\footnote{We do not capitalize ``student'' because it refers to the fact that William Gossett published his description of this distribution under the pen name ``a student.'' The distribution's name thus refers to the common noun rather than a capitalized proper noun.}} 
\label{app:multvar_t}

The multivariate student $t$-distribution extends the univariate student $t$-distribution to $d$-dimensional random vectors. 
It can be represented using random variables and vectors as:
\begin{equation} \label{eq:multvar_t_prop_rep}
    \vx = \vmu + {\mathbf y} \sqrt{\frac{\nu}{u}} \,,
\end{equation}
where the length-$d$ vector ${\mathbf y}$ is distributed as the multivariate normal $\mathcal{N}_d({{\mathbf y} \mid \mathbf 0},{\vSigma})$ with mean ${\mathbf 0}$ and covariance $\vSigma$, and the scalar $u$ has the chi-squared distribution $\chi^2 (u \mid \nu)$ with $\nu$ degrees of freedom.
The corresponding probability density function reads
\begin{align}
    t_\nu(\vx \mid \vmu, \vSigma) &= \frac{\Gamma\left[(\nu+d)/2\right]}{\Gamma(\nu/2)\nu^{d/2}\pi^{d/2}\left|{\vSigma}\right|^{1/2}} \notag \\
    & \quad \times \left[1+\frac{1}{\nu}(\vx-{\vmu})^{\intercal}{\vSigma}^{-1}(\vx-{\vmu})\right]^{-\frac{\nu+d}{2}} \,, \label{eq:multvar_t_pdf}
\end{align}
where the scale matrix $\vSigma$ is not to be confused with the distribution's covariance matrix, which happens to be
\begin{equation} \label{eq:t_cov_scale_mat_relation}
\vb{C} = \frac{\nu}{\nu-2} \vSigma \,,
\end{equation}
if $\nu > 2$ and is otherwise undefined. 
Instead, $\vSigma$ refers to the covariance matrix of the normal vector ${\mathbf y}$ in the probability representation~\eqref{eq:multvar_t_prop_rep} of $\vx$.
Mean and mode of the distribution $t_\nu(\vx \mid \vmu, \vSigma)$ are $\vmu$. 
In the bivariate case (\ie, $d=2$), one finds for the normalization factor in Eq.~\eqref{eq:multvar_t_pdf} to be
\begin{equation}
    \frac{\Gamma \left(\frac{\nu +2}{2}\right)}{\pi \, \nu \Gamma \left(\frac{\nu }{2}\right)}= \frac{1}{2\pi} \,.
\end{equation}

By inspecting the probability representation~\eqref{eq:multvar_t_prop_rep}, one sees that marginalizing $t_\nu(\vmu, \vSigma)$ over any subset of random variables $x_s$ results in $t_\nu(\vmu_s, \vSigma_s)$.
Here, $\vmu_s$ ($\vSigma_s)$ is obtained by dropping all the components in $\vmu$ (rows and columns in $\vSigma$) that are not associated with $x_s$.
This property of the multivariate student $t$-distribution is inherited from the multivariate normal distribution since $\vb{y}$ in Eq.~\eqref{eq:multvar_t_prop_rep} is a normally distributed vector. 
For $d=2$, it implies that the two marginal distributions of a bivariate $t$-distribution with $\nu$ degrees of freedom are univariate $t$-distributed with the same $\nu$.

\emph{Why is the multivariate $t$-distribution relevant to this work?}
Consider a normal vector with an unknown mean $\vmu$ and unknown covariance $\vSigma$, and assume that the pair $\vmu, \vSigma$ is NIW-distributed. 
For posterior (or even prior) prediction, one must understand the distribution of such a normal model integrated over its parameters' NIW distribution (see Eq.~\eqref{eq:posterior_pred_formal}). 
The integration is equivalent to stating that the randomness in the normal vector is independent of the randomness in the NIW distribution of its parameters. 
This independence is known as a mixture-model assumption, and the distribution obtained after mixing is known as the unconditional distribution of the model for the original vector. 
It turns out~\cite{gelman2013bayesian,pml2Book} that this unconditional distribution, which includes the uncertainty on the parameters, is the multivariate student $t$-distribution~\eqref{eq:posterior_pred}. 
%

\emph{Why is the posterior predictive distribution~\eqref{eq:posterior_pred} a multivariate student $t$-distribution?}
The probability representation~\eqref{eq:multvar_t_prop_rep} is one way to see why the construction of this mixture model
results in a multivariate $t$-distribution. 
To keep this explanation to a technical minimum while covering the conceptual issue at hand, we give the argument in the case $d=1$; 
the interested reader will check that the argument carries through in the multivariate case too. 
A normal variable $X \sim \mathcal{N}_1(\mu,\sigma^2)$ with mean $\mu$ and variance $\sigma^2$
can be constructed from a standard normal variable $Z \sim \mathcal{N}_1 (0,1)$ via the transform $X=\mu+\sigma Z$.
Next, if we now declare that, independently of the randomness in $Z$,  the scalar $\sigma^2$ should be inverse-chi-squared distributed, the resulting unconditional distribution of $X$ mixed with this inverse-chi-squared can be identified by replacing $\sigma^2$ by the reciprocal of a scaled chi-squared variable $\theta /u$, where $\theta$ is a positive constant. 
One then obtains immediately
\begin{equation}
    \text{unconditional}\; X=\mu+ Z \sqrt{\frac{\theta}{u}} \,,
\end{equation}
where $Z$ and $u$ are independent of each other. 
This is equivalent to the representation in Eq.~\eqref{eq:multvar_t_prop_rep} in the case $d=1$ (noting that the inverse-chi-squared distribution is the one-variable version of the inverse-Wishart distribution), proving that $X$ is univariate $t$-distributed with degrees of freedom $\nu$ given by those of $u$, and scale $\sigma=\sqrt{\theta/\nu}$. 
To use this constructive mixing procedure in the case $d>1$ to lead to the representation formula~\eqref{eq:multvar_t_prop_rep}, one then only needs to start by representing $\vb{X} \sim \mathcal{N}_d(\vmu,\vSigma)$ as 
\begin{equation}\label{eq:choleski_transform}
\vb{X} = \vb{R}\vb{Z}+ \vmu \,, 
\end{equation}
where $\vb{R}$ is the Cholesky factor of $\vSigma = \vb{R}\vb{R}^\intercal$. The details are omitted.  
Furthermore, the transform in Eq.~\eqref{eq:choleski_transform} allows one to draw numerous random samples from normal distributions with different mean vectors and covariance matrices computationally efficiently. 
It also allows for the efficient generation of $t$-distributed variables and vectors via the representation~\eqref{eq:multvar_t_prop_rep}.
\section{Confidence regions of the bivariate $t$-distribution} \label{app:conf_region_bivar_t}

We discuss in this Appendix how to compute confidence regions of the bivariate $t$-distribution ($d=2$)
\begin{equation} \label{eq:bivariate_t_distr}
    t_\nu(\vx \mid \vmu, \vSigma) = \frac{1}{2\pi \sqrt{|\vSigma|}}
    \left[1+\frac{(\vx-\vmu)^{\intercal}{\vSigma}^{-1}(\vx-\vmu)}{\nu}\right]^{-\frac{\nu+2}{2}}\!\!\!,
\end{equation}
with mean vector $\vmu$, $\nu$ degrees of freedom, and symmetric, positive-definite scale matrix $\vSigma$ and its inverse $\vSigma^{-1}$.
Our task is to determine a $(\alpha\times 100)\%$ confidence region $\Omega_\alpha$ for which
\begin{equation} \label{eq:CR_alpha}
    \alpha = \int_{\Omega_\alpha} \dd{\vx} t_\nu(\vx \mid \vmu, \vSigma) \,.
\end{equation}
Confidence regions are not unique in general; we focus here on those that are elliptical and centered at $\vmu$.

The analytic calculation of these confidence regions can be conveniently carried out in polar coordinates because the $t$-distribution~\eqref{eq:bivariate_t_distr} belongs to the class of elliptical distributions that depend on only the quadratic form
\begin{equation} \label{eq:mahalanobis}
    \rho^2 = (\vx-\vmu)^\intercal {\vSigma}^{-1} (\vx-\vmu) \,.
\end{equation}
%
The positive definiteness of $\vSigma$ (and $\vSigma^{-1}$) implies $\rho^2 >0$ and allows us to write 
\begin{equation}
    \vSigma^{-1} = \vR^\intercal \vR \,,
\end{equation}
with $\vR = \sqrt{\vLambda^{-1}} \vQ^\intercal $, the diagonal matrix $\vLambda$ containing the eigenvalues of $\vSigma$, and the matrix $\vQ$ having the corresponding orthonormal eigenvectors as columns.
In terms of the transformed coordinates $\vxbar = \vR (\vx - \vmu)$, one then finds that $\rho^2 = \vxbar^\intercal \vxbar$, such that 
\begin{equation}
\vx = \vR^{-1} \vxbar + \vmu \,, \quad \text{with} \quad
    \vxbar = \rho \begin{pmatrix}
    \cos \varphi \\ \sin \varphi
    \end{pmatrix} 
\end{equation}
and $\vR^{-1} = \vQ \sqrt{\vLambda} $, and the angle $\varphi = \operatorname{arctan2}(\vxbar_2, \vxbar_1)$. 
Here, $\operatorname{arctan2}(y,x)$ is the inverse-tangent function which returns the unique angle in the interval $(-\pi,\pi]$, corresponding to the ratio $y/x$, within its correct quadrant. 
The Jacobian determinant of this transformation reads
\begin{align}
\left| \frac{\partial(\vx_1, \vx_2)}{\partial(\rho, \varphi)}\right| 
= \begin{vmatrix}
      \frac{\partial \vx_1}{\partial \rho} & \frac{\partial \vx_1}{\partial \varphi} \\[2pt]
      \frac{\partial \vx_2}{\partial \rho} & \frac{\partial \vx_2}{\partial \varphi}
    \end{vmatrix} &= |\vR^{-1}| 
    \begin{vmatrix}
      \cos \varphi & -\rho \sin \varphi \\
      \sin \varphi & \rho \cos \varphi
    \end{vmatrix}\,, \notag \\
&= |\vR^{-1}| \, \rho = \sqrt{|\vSigma|} \, \rho \,. \label{eq:jacobian}
\end{align}
In the last step, we used that $|\vR^{-1}| = |\vQ||\sqrt{\vLambda}| = \sqrt{|\vSigma|}$.

We now look for a confidence region $\Omega_\alpha$ that has a constant radius in the new coordinate system, and spans the entire range $\varphi \in (-\pi,\pi]$. This means that, with the Jacobian determinant~\eqref{eq:jacobian}, we can now evaluate Eq.~\eqref{eq:CR_alpha} as follows:\footnote{Integrals of the form of Eq.~\eqref{eq:alpha_integral} can be solved using the integral representation of the Gauss hypergeometric function ${}_2F_1(a,b,c;z)$, which is, \eg, relevant for calculations at $d\neq 2$.}
\begin{align}
    \alpha &= \int_{\Omega_\alpha} \dd{\vx} t_\nu(\vx \mid \vmu, \vSigma) \,, \notag \\
    &= \frac{2\pi \sqrt{|\vSigma|}}{2\pi\sqrt{|\vSigma|}} \int_0^{\rho_0(\alpha)} \dd{\rho} \rho \left[1+\frac{\rho^2}{\nu}\right]^{-\frac{\nu+2}{2}} \,, \notag \\ 
    &=  1 - \left(\frac{\nu}{\nu + \rho_0^2(\alpha)}\right)^\frac{\nu}{2} \,. \label{eq:alpha_integral}
\end{align}
%
%
Hence, the ellipse encompassing the desired $(\alpha\times 100)\%$ confidence region, via the coordinate system given by the problem's eigenvectors, is parametrized as
\begin{align}
\vx(\varphi) &= \rho_0(\alpha) \, \vR^{-1}
\begin{pmatrix}
    \cos \varphi \\ \sin \varphi
    \end{pmatrix} + \vmu\,,
\intertext{with $\varphi \in (-\pi,\pi]$ and}
  \rho_0(\alpha) &= +\sqrt{\frac{\nu}{(1-\alpha)^\frac{2}{\nu}} - \nu} \,. \label{eq:rho_a_alpha} 
\end{align}
The major and minor axes have length $\rho_0 \max(\vLambda_{11}, \vLambda_{22})$ and $\rho_0 \min(\vLambda_{11}, \vLambda_{22})$, respectively, and are oriented in the directions of their corresponding eigenvectors.
The axis associated with $\vLambda_{ii}$ is rotated by the angle
\begin{equation}
    \theta = \arctan \frac{\vQ_{2i}}{\vQ_{1i}} \,,
\end{equation}
with respect to the $x$-axis.
From the quadratic form~\eqref{eq:mahalanobis}, one notes that $\rho^2 = d_\mathrm{MD}^2(\vx; \vmu, \vb{C}) \frac{\nu}{\nu-2}$, with the Mahalanobis distance (MD)
\begin{equation}
    d_\mathrm{MD}(\vx; \vmu, \vb{C}) = \sqrt{(\vx - \vmu)^\mathsf{T} \vb{C}^{-1} (\vx - \vmu)}
\end{equation}
and mean vector $\vmu$ and covariance matrix $\vb{C} = \frac{\nu}{\nu-2} \vSigma$ associated with the bivariate $t$-distribution~\eqref{eq:bivariate_t_distr}.
Hence, the ellipse encompassing the desired $(\alpha\times 100)\%$ confidence region corresponds to the contour with constant 
\begin{equation}
    d_\mathrm{MD}(\vx(\varphi); \vmu, \vb{C}
    ) 
    \equiv \rho_0(\alpha) \sqrt{\frac{\nu-2}{\nu}} \quad \text{for $\varphi \in (-\pi,\pi]$}  \,.
\end{equation}

Numerically, one can straightforwardly validate the obtained confidence regions by sampling the bivariate $t$-distribution~\eqref{eq:bivariate_t_distr}
and calculating the ratio of the number of samples that fall within the confidence region and the total number of samples, $\alpha \approx N_\text{in} / N_\text{tot}$.
The sampling points inside the confidence region, characterized by $(\alpha, \rho_0(\alpha))$, fulfill the inequality $\rho^2 \leqslant \rho_0^2(\alpha)$, where the left-hand side is given by Eq.~\eqref{eq:mahalanobis}.
We provide the Python function \texttt{plot\_confregion\_bivariate\_t($\ldots$)} in our GitHub repository~\cite{saturationGitHub} for plotting these confidence ellipses evaluated at one or multiple given values $\alpha$.

\emph{What confidence regions should one show in figures?}
We suggest showing the confidence regions associated with $\alpha = \{0.5, 0.8, 0.95, 0.99\}$, since these are common percentiles used in UQ studies. However, this choice is also a matter of preference and could depend on practical considerations such as how much uncertainty or tolerance exists for other aspects of a given study. For the use of a single percentile, we suggest $\alpha=0.95$ rather than higher or lower ones, since this level gives a good sense of the full support of the distribution, without considering events that are far out in its tails.

For completeness, let us study $\lim_{\nu \to \infty} P(\vx \mid \vmu, \vSigma) = \mathcal{N}_2(\vx \mid \vmu, \vSigma)$.
In this limit, we obtain a (bivariate) normal distribution and thus 
\begin{align}
    \lim_{\nu \to \infty} \rho_0(\alpha) \equiv \rho_{0,\infty}(\alpha) 
    &= \sqrt{-2\ln(1 - \alpha)} \,,
\intertext{which results in}
    \alpha(\rho_{0,\infty}) &= 1 - \exp[-\frac{\rho_{0,\infty}^2}{2}] \label{eq:normal_regions}\,,
\intertext{and thus, \eg,}
    \alpha(\rho_{0,\infty} = 1) &\approx 0.393469 \,, \\
    \alpha(\rho_{0,\infty} = 2) &\approx 0.864665 \,, \\
    \alpha(\rho_{0,\infty} = 3) &\approx 0.988891 \,, 
\end{align}
%
since Eq.~\eqref{eq:normal_regions} is identical to the one that determines the circular confidence regions for a standard bivariate normal distribution. 
One notes that these confidence levels differ from the well-known 68--95--99.7 rule for univariate normal distributions. 

Finally, we discuss that the outlined validation of the confidence ellipses can also be used to fit bivariate $t$-distributions $t_\nu(\vmu,\vSigma)$ to a set of given random samples drawn from an (unknown) distribution.
We provide the Python function \texttt{fit\_bivariate\_t(...)} for this task.
Since both the mean vector $\vmu$ and covariance matrix $\vb{C}$, which is proportional to the scale matrix $\vSigma$ (see Eq.~\eqref{eq:t_cov_scale_mat_relation}), can be straightforwardly estimated from the samples, it remains to optimize the degree of freedom~$\nu$.
To this end, for a given confidence level $\alpha$, one can use root finding (\eg, the bisection method) to find the integer $\nu$ in Eq.~\eqref{eq:rho_a_alpha} such that the given and estimated confidence levels match up to a desired tolerance, if possible.
As before, the estimated confidence level is given by the ratio of the number of samples that fall within the confidence region $N_\text{in}$ [\ie, those samples whose $\rho$ fulfils $\rho^2 \leqslant \rho_0^2(\alpha)$ for given $\nu$ and $ \vb{C}$] and the total number of samples $N_\text{tot}$.
We have found this simple strategy to be very efficient in practice.
Alternatively, one can determine $\nu$ by fitting the two marginals of the $t$-distribution using a Maximum Likelihood Estimation (MLE), as, \eg, implemented in \texttt{scipy.stats.rv\_continuous.fit}(...)~\cite{2020SciPy-NMeth}. 
Once the best-fit value for $\nu$ is known, the scale matrix is obtained via Eq.~\eqref{eq:t_cov_scale_mat_relation}.
This process can be repeated for different $\alpha$ to ensure the $t$-distribution fit is robust.




\section{Efficiency of the Bayesian model choices and sampling scheme}
\label{app:MC_for_BMM}

We discuss in this Appendix why the sampling method implemented in Sec.~\ref{sec:stat_framework_w_uncert} correctly samples from the posterior distributions described therein, why it is computationally efficient, and why our choices of prior and likelihood models are fruitful. 
For completeness, we justify here the procedure for sampling from the posterior of the pair of likelihood parameters $(\vmu,\vSigma)$ as well as from the posterior predictive distribution of $\mathbf{Y}$, though this paper focuses on an analysis of $\mathbf{Y}$'s posterior predictive distribution directly, bypassing the need to describe the posterior distribution of $(\vmu,\vSigma)$. 
Were it not for the explicitly computable nature of $\mathbf{Y}$'s posterior predictive ($t$-distribution), describing a sampling procedure for $(\vmu,\vSigma)$'s posterior would be required.
\subsection{Sampling scheme and its computational efficiency}

Recall the exact expression for the posterior density of our pair of parameters $(\vmu,\vSigma)$ from Eq.~\eqref{eq:full_posterior_unconditional}, 
\begin{multline} \
P(\vmu, \vSigma \mid \mods) = \int \dd{\vy} P(\Yc=\vy) \\
     \times \operatorname{NIW}(\vmu,\vSigma \mid  \kappa_n,\nu_n,\vmu_n,\vPsi_n; \mods, \Yc=\vy) \,. 
\end{multline}
By definition, this density is the mixture of the NIW distribution inside the integral on the right-hand side, with respect to the distribution of $\Yc=\{\Yci\}_{i=1}^n$. 
To sample from this mixture distribution, one only needs to sample a value $\vy$ from $\Yc$, and then sample from the NIW distribution inside the integral when the parameter $\vy$ is fixed. 

The same sampling method results in samples from the posterior predictive distribution, based on the expression in Eq.~\eqref{eq:full_posterior_pred}, where now one samples from the multivariate $t$-distribution after sampling a value $\vy$ from $\Yc$. 
To compute empirical statistics from such samples, the standard Monte Carlo method stipulates that a large number $Q$ of samples should be taken to  approximate the mathematical version of the statistic with the empirical version, \eg, approximating the posterior predictive mean of $\mathbf{Y}$ via
\begin{equation}
E_\text{post}[\mathbf{Y}] \approx \frac{1}{Q}\sum_{q=1}^Q \vb{y}'_{q}\,,
\end{equation} 
where the value $\vb{y}'_{q}$ is sampled from the $t$-distribution in the second line of Eq.~\eqref{eq:full_posterior_pred}, where, again, $\vy$ is taken to be the $q$th sample from $\Yc$.  

The sampling methodology we describe in Sec.~\ref{sec:stat_framework_w_uncert} and use in this work purports to implement the above with $Q=10^8$ samples, while choosing to sample the $t$-distribution $M=100$ times per sample from $\Yc$, instead of just once, thereby reducing the number of samples taken from $\Yc$ from $Q=10^8$ to $Q/M=N=10^6$.   
In other words, an outside loop samples from $\Yc$ only $N=10^6$ times, resulting in $N$ independent samples $\vb{y}_{q}$, for $~q=1,2,\ldots,N$, and for each $q$ from 1 to $N$, in an inside loop, we draw $M=100$ fresh independent samples from the distributions of interest, assuming $\vb{y}=\vb{y}_{q}$, using either the NIW distribution for $(\vmu,\vSigma)$ from the second line in Eq.~\eqref{eq:full_posterior_unconditional}, or the $t$ distribution from the prediction $\vb{y}'$ from the second line in Eq.~\eqref{eq:full_posterior_pred}. 

This procedure still results in $Q=N\times M=10^8$ independent samples for the posterior objects of interest, as in the classical method where $M=1$ and $Q=N=10^8$. 
However, reducing $N$ and increasing $M$ appears computationally more efficient in our setting. 
There are several reasons for this.

Drawing from the NIW distribution or the $t$-distribution is not computationally costly, so it is possible to draw a total of $Q=10^8$ $t$ or NIW samples. 
The overall uncertainty in our posteriors is attributed principally to the deviations between the various models, which are taken into account in the NIW and $t$ distributions, particularly because these distributions have heavy tails due to the small number of degrees of freedom in our posteriors. 
The overall uncertainty is only secondarily due to individual model uncertainty reports $P(\Yci)$, which have light tails and are of a smaller magnitude. 
Therefore, more computational effort should be spent exploring the NIW and $t$-distributions (total of $Q=10^8$ samples) in our mixed posteriors, and less needs to be spent on exploring individual model uncertainties (only $N=10^6$ samples from the $\Yci$'s). 

On the other hand, there is a trade-off between how much smaller $N$ should be compared with $Q$, \ie, how much greater $M$ should be than~1. 
Indeed, the precision in the Monte Carlo method is well known to be inversely proportional to the square root of the number of samples. 
Therefore, since part of the error in our Monte Carlo scheme is attributed to the use of $N$ samples from $\Yc$, one expects an error of order $1/\sqrt{N}=10^{-3}$, which is acceptable. 
The Monte Carlo errors are also proportional to the scale of uncertainty in the statistics being approximated. 
This scale is measured using standard deviations. 
As mentioned in the previous paragraph, those standard deviations are significantly larger for our posterior NIW and $t$ distributions for fixed $\vb{y}$ than for our distributions $P(\Yci)$. 
Thus, when dealing with the Monte Carlo error for the NIW and $t$ samples, which is proportional to $1/\sqrt{Q}=10^{-4}$, one must also keep in mind that the error's magnitude is also proportional to the corresponding standard deviations, which might legitimately be close to an order of magnitude larger than the standard deviations of the $P(\Yci)$, as can be seen, \eg, in Fig.~\ref{fig:satpoint_constraints}. 

This point explains why choosing $N=10^6$ and $Q=10^8$ results in a good precision, \ie, good convergence of Monte Carlo methods based on our sampling scheme with a good balance between the areas with higher uncertainty and those with lower uncertainty. 
In particular, the increase from $M=1$ to $M=100$ can be considered as a so-called \emph{variance reduction} scheme, in Monte Carlo parlance, since it takes advantage of the lower uncertainty from one section of our models by drawing fewer samples, compensating the higher levels of uncertainty elsewhere in the models with full numbers of samples. 

These ideas can be made more precise mathematically by using a classical formula for the unconditional variance of a mixture distribution, which is equal to the expected value of the conditional variance plus the variance of the conditional expectation, where one conditions by the mixing variable, in this case, the $\Yc$. 
One quantitative element that transpires in such details is the variance of empirical variances, which draws on a scale parameter related to the fourth central moment. 
It should be noted that for the $t$-distribution, the fourth central moment does not exist for degrees of freedom less than~5. 
Our fitted posterior $t$ distribution's degrees of freedom is equal to $9$ (see the text above Eq.~\eqref{eq:fit_bivar_t_setA}), whereas the number of degrees of freedom before mixing was $\nu_n=\nu_0+n-d+1=17$, thus we are in safe territory. However this points to our variance reduction scheme not being expected to work well with a significantly lower number $n$. 

While the mathematical details are omitted for conciseness, we offer here a simple analysis to show why the choice $M>1$ is a variance reduction technique in the case where, as we have with our setting, the discrepancy across models is significantly greater than the uncertainty within each model. 
The argument below refers to the $t$ distribution to help with readability, but is entirely general. 
Our Monte Carlo method proposes samples for the posterior predictive distribution of $\mathbf{Y}'=\mathbf{Y}'(\vy)$, where $\vy$ is distributed according to the developer-reported UQ, independently of the ($t$-distributed) posterior prediction for $\mathbf{Y}'(\vy)$ when $\vy$ is fixed. 
Now, for fixed $\vy$, we note that the expectation and variance $\mathbf{E}\left[\mathbf{Y}'(\vy)\right]$ and  $\mathbf{Var}[\mathbf{Y}'(\vy)]$ can also be considered as random variables since they are functions of the fundamentally uncertain $\vy$. 
We denote by $\mathbf{E}_{\vy}$ and $\mathbf{Var}_{\vy}$ the expectation and variance operators with respect to the randomness of $\vy$, which is specified by the developers' UQ. 
Now, let
\begin{align}
    \sigma_t^2 &:= \mathbf{E}_{\vy} [\mathbf{Var}[\mathbf{Y}'(\vy)]] \,, \\
 \sigma_{\vy}^2 &:=  
\mathbf{Var}_{\vy} [\mathbf{E}[\mathbf{Y}'(\vy)]] \,.
\end{align}
A straightforward calculation based on the aforementioned formula for the unconditional variance of a mixture leads to the following expression for the empirical mean and variance of all our $Q$ samples for $\mathbf{Y}'$ in our Monte Carlo method:
\begin{equation} \label{See_five}
\operatorname{Var}\left[\frac{1}{Q}\sum_{q=1}^Q\mathbf{Y}'_q\right] = \frac{\sigma_t^2}{Q}+\frac{\sigma_{\vy}^2}{N}
= \frac{1}{N}\left(\frac{\sigma_t^2}{M}+\sigma_{\vy}^2\right) \,.
\end{equation}
We must now realize that the limiting factor in ensuring an adequate number of Monte Carlo samples is the value $N$, since our method only takes $N$ samples from the developer UQ, while it produces $Q$ samples from the $t$ distribution.
Hence the variance in this Monte Carlo method with mixture can be considered as the last factor on the right-hand side of Eq.~\eqref{See_five}, namely $\left(\sigma_t^2/M+\sigma_{\vy}^2\right)$. Clearly, when $\sigma_t^2$ is significantly larger than $\sigma_{\vy}^2$, it is beneficial to use $M$ significantly larger than $1$ to reduce this overall variance. 
To produce a balanced Monte Carlo method, it is then judicious to choose the two terms in this variance expression to be equal to each other. 
This leads immediately to 
\begin{equation}
M = \frac{\sigma_t^2}{\sigma_{\vy}^2} \,.
\end{equation} 
In our setting, one can give Fig.~\ref{fig:satpoint_constraints} a visual inspection to see that, indeed, the discrepancy among models, which is represented by $\sigma_{t}$, exceeds by a significant factor the intra-model uncertainties, for which $\sigma_{\vy}$ is a summary measurement. 
The square of this factor would be a good selection for $M$.  

\subsection{Efficiency of prior and likelihood labels}

Let us explain in qualitative terms, as announced in Sec.~\ref{sec:stat_framework_model}, why our mapping of priors and likelihoods is efficient.
In a hierarchical model, there is not necessarily a single way to map response variables, models, and parameters to priors and likelihoods. 
We restrict the discussion here to two-level hierarchies, as is the relevant case for our study. 

To understand the ambiguity, consider, for instance, the base level of a two-level hierarchy. 
Though it is a probability model, it may be interpreted as a prior. 
This can, \eg, be the case if the base level refers to data that is external to the data used to inform the top-level model. 
But that is not the case for us: 
the base level~\eqref{eq:hmodel_base_level} uses the only data we have, namely all the $\YObsi$'s reported via their associated uncertainties $\Yci$. 
The base level can also be interpreted as a prior if the top-level model involves a latent variable, which one might wish to reconstruct. 
A Bayesian methodology is well-suited to reconstructing latent variables when they are part of the likelihood models. 
A na\"ive interpretation of a Bayesian hierarchical model would then hesitate over which of the two levels should be the sole likelihood equation, since both contain the latent variables, in our case, the $\YObsi$'s. 
Another interpretation could conclude that both levels of the hierarchy should be deemed the likelihood model, with priors being reserved for model parameters that are not interpreted as latent variables. 
This latter interpretation is what we do here. 
It avoids the conundrum of involving latent variables in the prior, but there is a stronger reason for this choice than latent variables. 

In the present study, the basic object, whose posterior distribution we are after, even before discussing posterior prediction, is solely the pair of parameters $(\vmu,\vSigma)$. 
Therefore, the top-level Eq.~\eqref{eq:hmodel_top_level}, which is based on that pair of parameters, should be part of the likelihood model, and any models for that pair $(\vmu,\vSigma)$ should be part of the prior. 
Since the variables $\Yci$ are mixing variables, which means by definition (\ie, by our original modeling choice) that their distribution is not influenced by the data, and the Bayesian analysis will not change their distribution, the most efficient choice is to label the base level Eq.~\eqref{eq:hmodel_base_level} as also being part of the likelihood. 
We think this labeling will help readers keep track of what part of the model is formed of parameters that need to be estimated using Bayes and which part of the model will not be affected by the Bayesian analysis. 
However, the base level's designation as part of the likelihood is, to some extent, immaterial, precisely since the $\Yci$'s are mixing variables whose laws are unaffected by the data and its analysis. 
In particular, the $\Yci$'s cannot be reconstructed more precisely than their original distributions.

\section{Additional Results}
\label{app:suppl_results}

This Appendix provides additional results obtained with the alternative prior Set~B. 
We also discuss how the results change if the functional SQMC700 is removed from the Bayesian analysis carried out in Sec.~\ref{sec:results_satbox_analysis}.

\begin{figure*}[tb]
    \centering
    \includegraphics[width=0.49\linewidth,page=3]{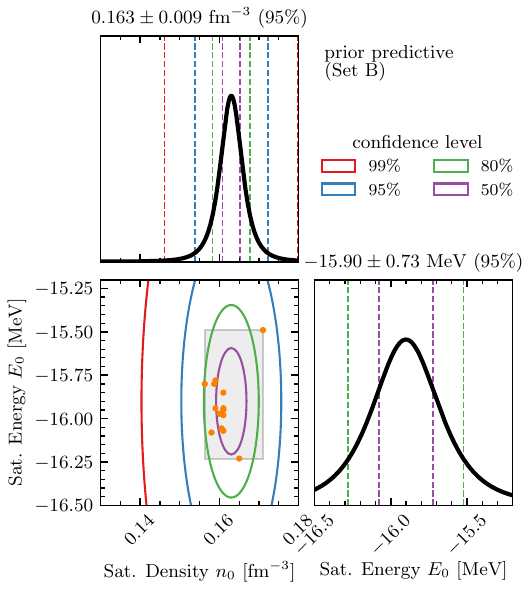}
    \includegraphics[width=0.49\linewidth,page=4]{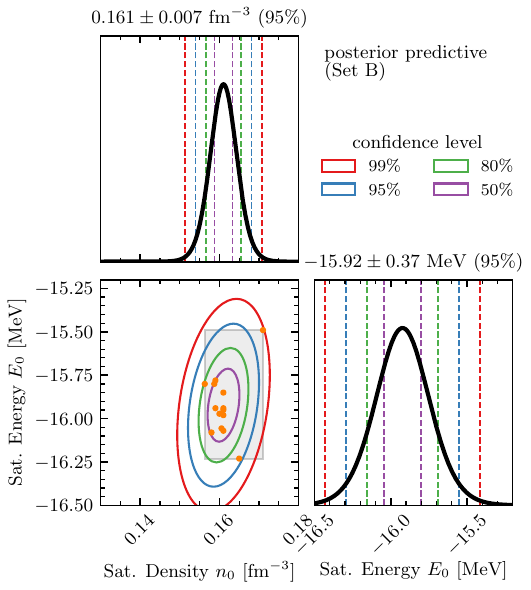}
    \caption{%
    Similar to Fig.~\ref{fig:satbox_analysis_set_A} but for the more weakly informed prior Set~B.
    } \label{fig:satbox_analysis_set_B}
\end{figure*}

\begin{figure}[tb]
    \centering
    \includegraphics[width=\linewidth, page=2]{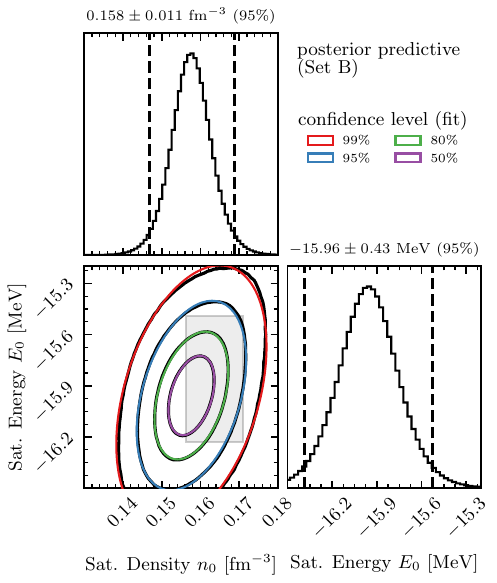}
    \caption{%
    Similar to Fig.~\ref{fig:full_analysis_setA} but for the more weakly prior Set~B.
    Note that the left panel in Fig.~\ref{fig:satbox_analysis_set_B} shows the associated (data-agnostic) prior predictive.
    See the main text for details.
    } \label{fig:full_analysis_setB}
\end{figure}

\begin{figure*}[tb]
    \centering
    \includegraphics[width=0.49\linewidth, page=4]{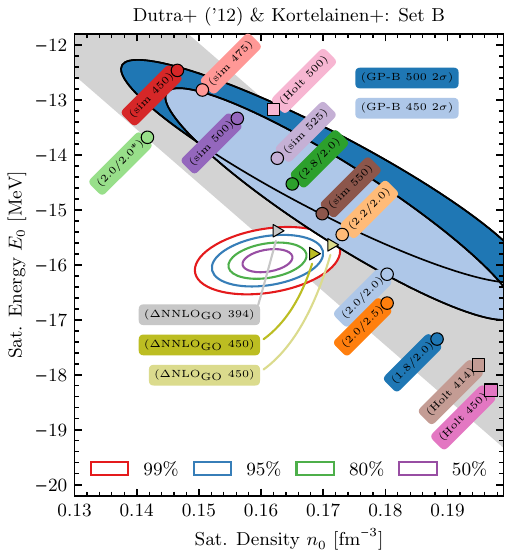}
    \includegraphics[width=0.49\linewidth, page=6]{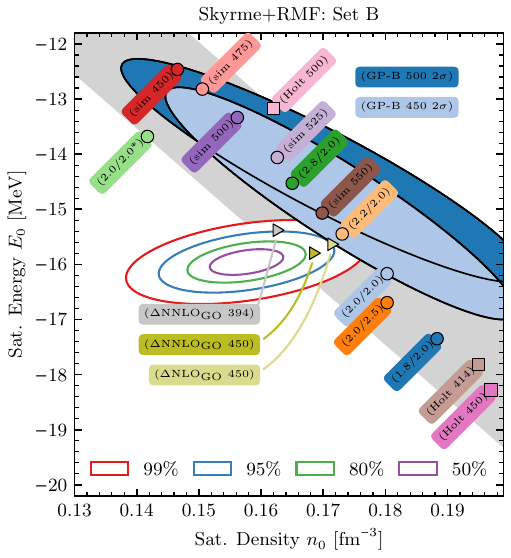}
    \caption{%
    Similar to Fig.~\ref{fig:eft_satpoints} but for the more weakly prior Set~B.
    } \label{fig:eft_satpoints_setB}
\end{figure*}

\begin{figure*}[tb]
    \centering
    \includegraphics[width=0.49\linewidth, page=4]{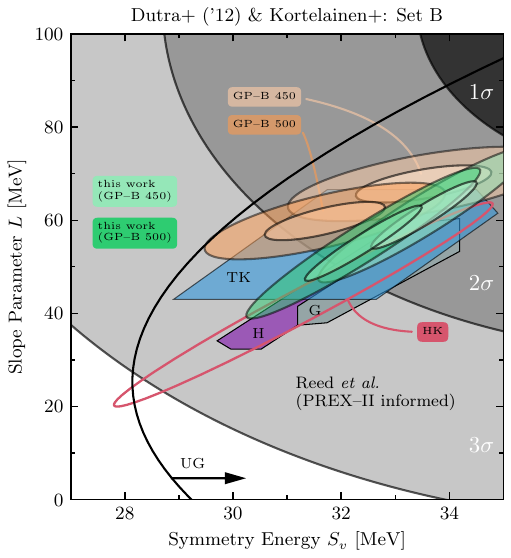}
    \includegraphics[width=0.49\linewidth, page=6]{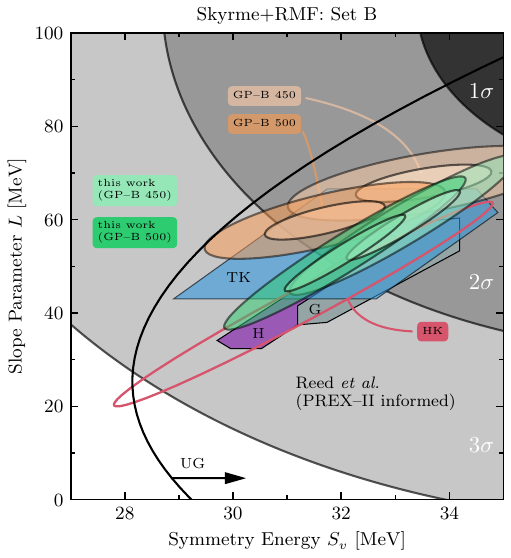}
    \caption{%
    Similar to Fig.~\ref{fig:esym} but for the more weakly prior Set~B.
    } \label{fig:esym_setB}
\end{figure*}

Figure~\ref{fig:satbox_analysis_set_B} shows the results of the analysis for the data points that define the saturation box~\eqref{eq:satbox}.
For Set~B, we obtain the NIW posterior~\eqref{eq:posterior} with the updated hyperparameters $(\kappa_n = 15,\nu_n = 18)$, and
\begin{equation} 
    \vmu_n^{\text{(B)}} \approx \begin{bmatrix}
    0.161 \\ -15.92
    \end{bmatrix} \,, \quad
    \vPsi_n^{\text{(B)}} \approx \mqty[0.013^2 & 0.051^2 \\ 0.051^2 & 0.70^2] \,.
\end{equation}
We also obtain the data-informed posterior predictive~\eqref{eq:posterior_pred} (right panel) is given by the $t_\nu(\vmu, \vPsi)$ with hyperparameters:
\begin{equation} 
\nu^{\text{(B)}} = 17 \,, \quad 
    \vmu^{\text{(B)}} = \vmu_n^{\text{(B)}} \,, \quad
    \vPsi^{\text{(B)}} \approx \mqty[0.003^2 & 0.013^2 \\ 0.013^2 & 0.18^2] \,.
\end{equation}
At the 95\% credibility level, we therefore obtain the two marginal distributions:
\begin{subequations} 
\begin{align}
    n_0^{\text{(B)}} &\approx 0.161 \pm 0.007 \fmiq \,,\\ 
    E_0^{\text{(B)}} &\approx -15.92 \pm 0.37 \MeV \,.
\end{align}
\end{subequations}

Figure~\ref{fig:full_analysis_setB} shows the results of the analysis with both Skyrme and RMF models considered, as in Sec.~\ref{sec:results_full_analysis}. 
The fitted posterior predictive $t_\nu(\vmu, \vPsi)$ has $\nu_n = 9 \ll \infty$ and
\begin{equation} 
    \vmu_n^{\text{(B)}} \approx \begin{bmatrix}
        0.158 \\ -15.96
    \end{bmatrix} \,, \quad
    \vPsi_n^{\text{(B)}} \approx \mqty[0.005^2 & 0.019^2 \\ 0.019^2 & 0.19^2] \,.
\end{equation}

Furthermore, Figs.~\ref{fig:eft_satpoints_setB} and~\ref{fig:esym_setB} present our results for benchmarking chiral EFT interactions and comparing the nuclear symmetry energy, respectively.
For the results in Sec.~\ref{sec:results_satbox_analysis} with Set~B and ``GP--B 500,'' we find
\begin{equation} 
    \vmu^{\text{(A)}} 
    \approx \begin{bmatrix}
    32.4 \\ 55.1
    \end{bmatrix} \,,  \quad 
    \vSigma^{\text{(A)}} \approx \mqty[1.1^2 & 2.9^2 \\ 2.9^2 & 8.1^2] \,,
\end{equation}
and for ``GP--B 450''
\begin{equation} 
    \vmu^{\text{(A)}} \approx \begin{bmatrix}
    33.5 \\ 61.1
    \end{bmatrix} \,, \quad 
    \vSigma^{\text{(A)}} \approx \mqty[1.0^2 & 2.6^2 \\ 2.6^2 & 7.3^2] \,,
\end{equation}
Likewise, for the results in Sec.~\ref{sec:results_full_analysis} with Set~B and ``GP--B 500,'' we find
\begin{equation} 
    \vmu^{\text{(B)}} 
    \approx \begin{bmatrix}
    32.1 \\ 52.8
    \end{bmatrix} \,,  \quad 
    \vSigma^{\text{(B)}} \approx \mqty[1.1^2 & 3.0^2 \\ 3.0^2 & 8.2^2] \,,
\end{equation}
and for ``GP--B 450''
\begin{equation} 
    \vmu^{\text{(B)}} \approx \begin{bmatrix}
    33.2 \\ 58.9
    \end{bmatrix} \,, \quad 
    \vSigma^{\text{(B)}} \approx \mqty[1.1^2 & 2.7^2 \\ 2.7^2 & 7.5^2] \,.
\end{equation}
%

\begin{figure*} 
    \centering
    \includegraphics[width=0.49\linewidth,page=1]{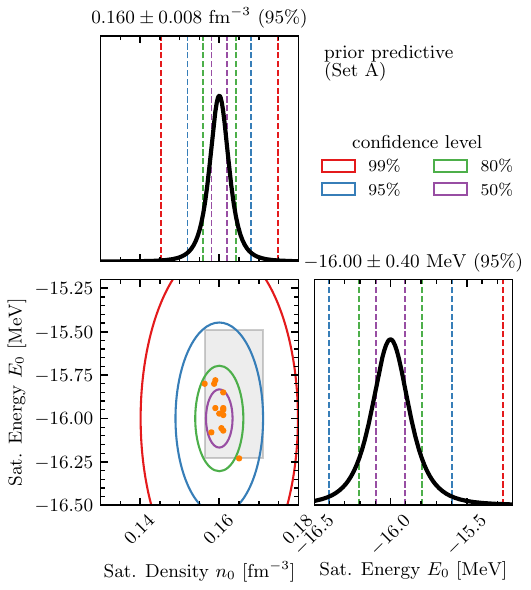}
    \includegraphics[width=0.49\linewidth,page=2]{figures/corner_satbox_wo_SQMC700.pdf}
    \caption{%
    Similar to Fig.~\ref{fig:satbox_analysis_set_A} (prior Set~A) but ignoring the Skyrme functional SQMC700, which sets the upper-right corner of the saturation box constructed in Ref.~\cite{Drischler:2015eba}.
    } \label{fig:satbox_analysis_wo_SQMC700_set_A}
\end{figure*}
\begin{figure*} 
    \centering
    \includegraphics[width=0.49\linewidth,page=3]{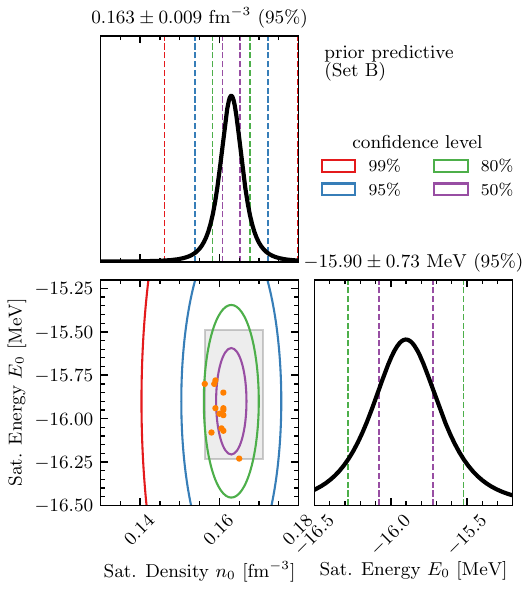}
    \includegraphics[width=0.49\linewidth,page=4]{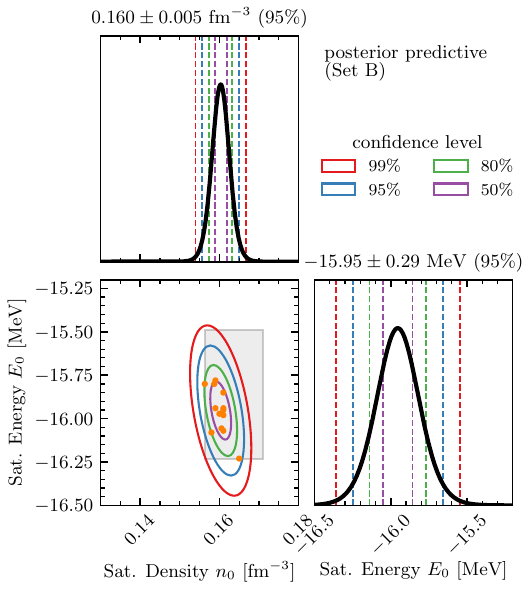}
    \caption{%
    Similar to Fig.~\ref{fig:satbox_analysis_wo_SQMC700_set_A} but for prior Set~B. Notice that the data point that sets the upper-right corner of the saturation box constructed in Ref.~\cite{Drischler:2015eba} is ignored.
    } \label{fig:satbox_analysis_wo_SQMC700_set_B}
\end{figure*}

\rev{Figures~\ref{fig:satbox_analysis_wo_SQMC700_set_A}} and~\ref{fig:satbox_analysis_wo_SQMC700_set_B} show the results of our analysis when the functional SQMC700 is removed from the analyzed dataset, leaving $n=13$ data points.
For Set~A, the data-informed posterior predictive~\eqref{eq:posterior_pred} (right panel) is given by the $t_\nu(\vmu, \vPsi)$ with $\nu^{\text{(A)}} = 16$ and:
\begin{equation} 
    \vmu_n^{\text{(A)}} \approx \begin{bmatrix}
    0.160 \\ -15.96
    \end{bmatrix} \,, \quad
    \vPsi^{\text{(A)}} \approx \mqty[0.002^2 & -0.012^2 \\ -0.012^2 & 0.12^2] \,.
\end{equation}
At the 95\% credibility level, we therefore obtain the two marginal distributions:
\begin{subequations} \label{eq:marginal_full_wo_SQMC700_setA}
\begin{align}
    n_0^{\text{(A)}} &\approx 0.160 \pm 0.004 \fmiq \,,\\ 
    E_0^{\text{(A)}} &\approx -15.96 \pm 0.26 \MeV \,.
\end{align}
\end{subequations}
For Set~B, the data-informed posterior predictive~\eqref{eq:posterior_pred} (right panel) is given by the $t_\nu(\vmu, \vPsi)$ with $\nu^{\text{(B)}} = 16$ and
\begin{equation} 
    \vmu_n^{\text{(B)}} \approx \begin{bmatrix}
    0.160 \\ -15.95
    \end{bmatrix} \,, \quad
    \vPsi^{\text{(B)}} \approx \mqty[0.002^2 & -0.012^2 \\ -0.012^2 & 0.14^2] \,.
\end{equation}
At the 95\% credibility level, we obtain:
\begin{subequations} \label{eq:marginal_full_wo_SQMC700_setB}
\begin{align}
    n_0^{\text{(B)}} &\approx 0.160 \pm 0.005 \fmiq \,,\\ 
    E_0^{\text{(B)}} &\approx -15.95 \pm 0.29 \MeV \,,
\end{align}
\end{subequations}
which is statistically consistent with the results for Set~A given in Eq.~\eqref{eq:marginal_full_wo_SQMC700_setA} and those obtained in Sec.~\ref{sec:results_satbox_analysis} but (as expected) more tightly constrained.
Removing SQMC700 renders $(n_0,E_0)$ anti-correlated, with the Pearson correlation coefficient $\rho \approx -0.57$ (intermediately correlated) for Set~A and $\rho \approx -0.44$ (weakly correlated) for Set~B.
To be more conservative in our inference, we will keep SQMC700 in the data collection and treat it as an outlier at the prior level.

\begin{table*}[tb]
\renewcommand{\arraystretch}{1.2}
\caption{%
\rev{Details of the additional DFT constraints similar to Table~\ref{tab:constraints}.
All constraints are given as empirical distributions without additional UQ, with the number of point estimates (or random samples) denoted by $n_i$.}}
\label{tab:constraints_extended}
\begin{ruledtabular}
\begin{tabular}{lp{4.7cm}lp{6.9cm}}
Label & Type & $n_i$ & Comments\\ 
\colrule
Baldo+ ('12)~\cite{Baldo:2012hw} & Barcelona--Catania--Paris--Madrid (BCPM) & 1 & largely based on \emph{ab~initio} calculations of the symmetric and pure neutron matter EOSs with a small set of parameters; see Table~II in Ref.~\cite{Baldo:2012hw} \\
Sellahewa+ ('14)~\cite{Sellahewa:2014nia} & Gogny & 10 & comprehensive analysis of 11 Gogny interactions with emphasis on the isovector sector; these models are D1, D1S, D1N, D1M, D1P, D1AS, D250, D260, D280, D300, and GT2; see Table~I~\cite{Sellahewa:2014nia}; D1 and D1AS have the same saturation point so one of them is excluded \\
Bulgac+ ('17)~\cite{Bulgac:2017bho} & Skyrme-like minimal nuclear energy density functional (NEDF) & 1 & the functional SeaLL1 has a minimal set of parameters, constrained by the EOS calculations of pure neutron matter with chiral NN and 3N interactions, and contains terms not used in Skyrme models; see Table~III in Ref.~\cite{Bulgac:2017bho} \\
Bollapragada+ ('20)~\cite{Bollapragada:2020end} & Fayans & 1 & obtained using a comprehensive optimization analysis; see Table~III in Ref.~\cite{Bollapragada:2020end}; we consider only the parametrization with the lowest objective function value ($\mathbf{x}_1$) as the other parametrization ($\mathbf{x}_5$) results in a similar saturation point
\end{tabular}
\end{ruledtabular}
\end{table*}

\begin{figure} 
    \centering
    \includegraphics[width=\linewidth,page=13]{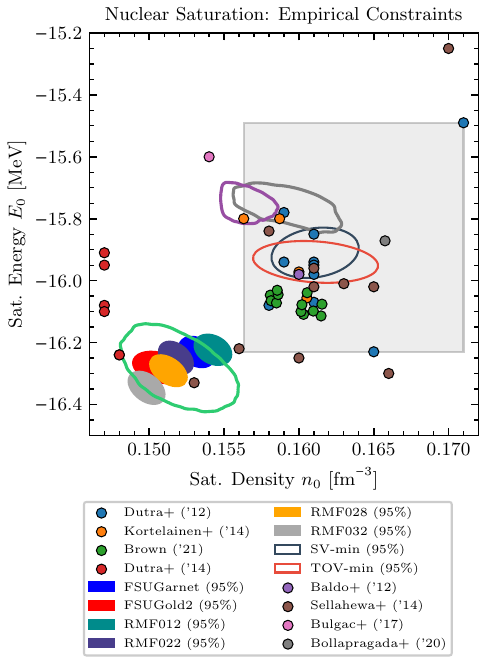}
    \caption{%
    \rev{Similar to Fig.~\ref{fig:satpoint_constraints} but including the 13 additional point-like DFT constraints detailed in Table~\ref{tab:constraints_extended}.
    }} \label{fig:satpoint_constraints_extended}
\end{figure}

\begin{figure*} 
    \centering
    \includegraphics[width=0.49\linewidth,page=2]{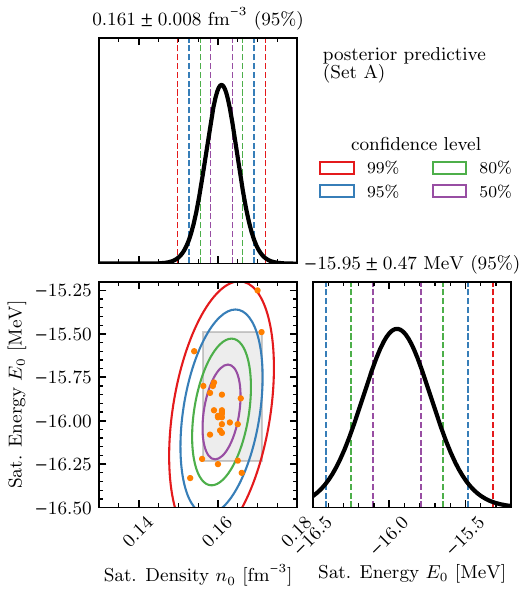}
    \includegraphics[width=0.49\linewidth,page=4]{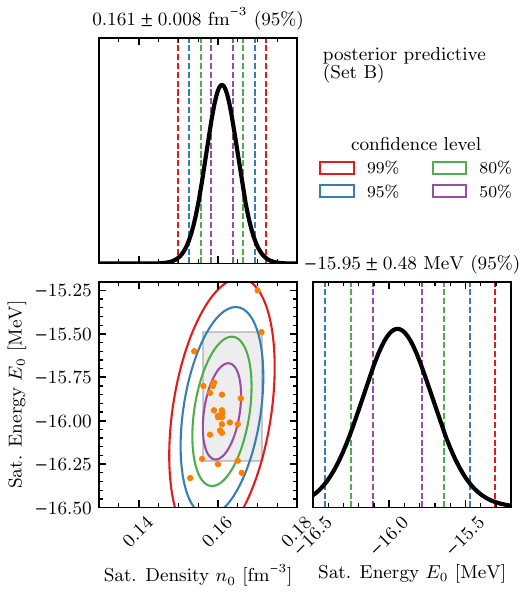}
    \caption{\rev{Similar to the right panels in Figs.~\ref{fig:satbox_analysis_set_A} and~\ref{fig:satbox_analysis_set_B} for the two prior sets but including the additional DFT constraints detailed in Table~\ref{tab:constraints_extended}. Note that the corresponding prior predictive distributions in the left panels in Figs.~\ref{fig:satbox_analysis_set_A} and~\ref{fig:satbox_analysis_set_B}, respectively, would remain unchanged, are thus not shown.}} \label{fig:satbox_analysis_extended}
\end{figure*}

\rev{Finally, we extended our analysis of the saturation box in Sec.~\ref{sec:results_satbox_analysis} to also consider the 13 point-like DFT constraints detailed in Table~\ref{tab:constraints_extended}, including Gogny interactions and a Fayans functional. 
The total number of DFT constraints considered in this particular analysis thus increases to $n=27$. 
Figure~\ref{fig:satpoint_constraints_extended} shows these constraints together with the ones in Table~\ref{tab:constraints} similar to Fig.~\ref{fig:satpoint_constraints}.
Five of these 13 functionals predict saturation points outside the original saturation box (gray box), with D1P having the highest ground-state energy. 

Figure~\ref{fig:satbox_analysis_extended} shows the posterior predictive distributions for the two prior Set~A and Set~B (see Eq.~\eqref{eq:posterior_setA} and Eq.~\eqref{eq:posterior_setB}), based on these $n=27$ DFT data points similar to the right panels in Figs.~\ref{fig:satbox_analysis_set_A} and~\ref{fig:satbox_analysis_set_B}.
The corresponding prior predictive distributions are shown in the left panels in Figs.~\ref{fig:satbox_analysis_set_A} and~\ref{fig:satbox_analysis_set_B}, respectively.
For prior Set~A, the posterior predictive (left panel in Fig.~\ref{fig:satbox_analysis_extended}) is given by the $t_\nu(\vmu, \vPsi)$ with $\nu^{\text{(A)}} = 30$ and:
\begin{equation} 
    \vmu_n^{\text{(A)}} \approx \begin{bmatrix}
    0.161 \\ -15.95
    \end{bmatrix} \,, \quad
    \vPsi^{\text{(A)}} \approx \mqty[0.004^2 & 0.018^2 \\ 0.018^2 & 0.23^2] \,.
\end{equation}
At the 95\% credibility level, for Set~A, we therefore obtain the following credible intervals for the two marginal distributions:
\begin{subequations}  
\begin{align}
    n_0^{\text{(A)}} &\approx 0.161 \pm 0.008 \fmiq \,,\\ 
    E_0^{\text{(A)}} &\approx -15.95 \pm 0.47 \MeV \,.
\end{align}
\end{subequations}
Similarly, for Set~B, the posterior predictive (left panel in Fig.~\ref{fig:satbox_analysis_extended}) is given by the $t_\nu(\vmu, \vPsi)$ with $\nu^{\text{(B)}} = 30$ and
\begin{equation} 
    \vmu_n^{\text{(B)}} \approx \begin{bmatrix}
    0.161 \\ -15.95
    \end{bmatrix} \,, \quad
    \vPsi^{\text{(B)}} \approx \mqty[0.004^2 & 0.018^2 \\ 0.018^2 & 0.23^2] \,.
\end{equation}
At the 95\% credibility level, we obtain:
\begin{subequations} 
\begin{align}
    n_0^{\text{(B)}} &\approx 0.161 \pm 0.008 \fmiq \,,\\ 
    E_0^{\text{(B)}} &\approx -15.95 \pm 0.48 \MeV \,,
\end{align}
\end{subequations}
which is statistically consistent with the corresponding results obtained in Sec.~\ref{sec:results_satbox_analysis} without the constraints in Table~\ref{tab:constraints_extended}: the credible intervals with the additional constraints contain those without the additional constraints. With those extended constraints, uncertainties become slightly larger, particularly in $E_0$, as expected from Fig.~\ref{fig:satpoint_constraints_extended}.
}

\bibliographystyle{apsrev4-2}
\bibliography{bayesian_refs,more_refs,bib}

\end{document}